# Μοντελοποίηση του οδικού διαπεριφερειακού δικτύου της Ελλάδας με χρήση ανάλυσης σύνθετων δικτύων (complex network analysis)

**Δημήτριος Τσιώτας**
Τμήμα Μηχανικών Χωροταξίας Πολεοδομίας και Περιφερειακής Ανάπτυξης,
Πανεπιστήμιο Θεσσαλίας, Πεδίον Άρεως, Βόλος, 38 334,
Τηλ +30 24210 74446, fax: +302421074493
E-mails: spolyzos@uth.gr; tsiotas@uth.gr

**Περίληψη**

Αυτό το άρθρο μελετά το ελληνικό διαπεριφερειακό δίκτυο οδικών μεταφορών (Greek road network - GRN) εφαρμόζοντας ανάλυση σύνθετων δικτύων (complex network analysis - CNA) και εμπειρική προσέγγιση. Η μελέτη αποσκοπεί στην εξόρυξη της κοινωνικοοικονομικής πληροφορίας που είναι ενσωματωμένη στην τοπολογία του GRN και να ερμηνεύσει τον τρόπο με τον οποίο το δίκτυο αυτό υπηρετεί και προάγει την περιφερειακή ανάπτυξη. Στην ανάλυση που πραγματοποιείται προκύπτει ότι η τοπολογία του GRN υπόκειται στην επίδραση των χωρικών περιορισμών, η οποία προκύπτει συγγενική με το θεωρητικό πρότυπο του *δικτυώματος* (lattice network). Επίσης, διαφαίνεται ότι η δομή του δικτύου περιγράφεται από *βαρυτική λειτουργία*, έχοντας προτεραιότητα την εξυπηρέτηση των περιοχών ανάλογα με τον πληθυσμό τους, και ότι η σύνδεσή του με περιφερειακά μεγέθη σκιαγραφεί το στοιχειώδες πρότυπο της «*ανάπτυξης εκεί που υπάρχει δρόμος*». Από μελέτη προκύπτουν ενδιαφέρουσες αντιθέσεις μεταξύ των μητροπολιτικών και μη-μητροπολιτικών (εξαιρώντας την Αττική και Θεσσαλονίκη) θεωρήσεων. Συνολικά, το άρθρο αναδεικνύει την αποτελεσματικότητα της χρήσης της ανάλυσης των σύνθετων δικτύων στη μοντελοποίηση των χωρικών δικτύων και ειδικότερα των συστημάτων μεταφορών και επιδιώκει την προώθηση χρήσης του παραδείγματος των δικτύων στις χωρικές και περιφερειακές εφαρμογές.

**Λέξεις κλειδιά:** πολύπλοκα δίκτυα, χωρικά δίκτυα, επιστήμη των δικτύων, αναγνώριση προτύπων, οδικές μεταφορές, περιφερειακή ανάπτυξη.

**Abstract**

This article studies the interregional Greek road network (GRN) by applying complex network analysis (CNA) and an empirical approach. The study aims to extract the socioeconomic information immanent to the GRN's topology and to interpret the way in which this road network serves and promotes the regional development. The analysis shows that the topology of the GRN is submitted to spatial constraints, having lattice-like characteristics. Also, the GRN's structure is described by a gravity pattern, where places of higher population enjoy greater functionality, and its interpretation in regional terms illustrates the elementary pattern expressed by "regional development through road construction". The study also reveals some interesting contradictions between the metropolitan and non-metropolitan (excluding Attica and Thessaloniki) comparison. Overall, the article highlights the effectiveness of using complex network analysis in the modeling of spatial networks and in particular of transportation systems and promotes the use of the network paradigm in the spatial and regional research.

**Keywords:** complex networks, spatial networks, network science, pattern recognition, regional development.

**1. Εισαγωγή**

Τα οδικά δίκτυα αποτελούν τα πιο διαδεδομένα και προσβάσιμα δίκτυα χερσαίων μεταφορών, λόγω της επικράτησης του αυτοκινήτου ως μέσο ιδιωτικής μετακίνησης (Kurant Thiran, 2006; Πολύζος, 2011; Barthelemy, 2011; Polyzos et al., 2014). Ο όρος «χερσαία δίκτυα μεταφορών» αναφέρεται στα δίκτυα που αναπτύσσονται στην ξηρά και εξυπηρετούν τη διενέργεια των μεταφορών, δίχως την παρεμβολή θαλασσίων ή εναερίων μέσων. Λαμβάνοντας υπόψη ότι οι μεταφορές μπορούν να θεωρηθούν ως πτυχή της ανθρώπινης επικοινωνίας που υποβάλλεται σε αναπόφευκτους χωρικούς περιορισμούς (Rodrigue et al., 2013, Tsiotas and Polyzos, 2015), καθίσταται προφανές ότι η δομή και η μορφή που αποκτούν διαχρονικά αυτού του είδους τα δίκτυα αντανακλούν τις εκάστοτε ιστορικές και κοινωνικοοικονομικές ανάγκες για ανθρώπινη επικοινωνία, αλλά και εξαρτώνται από τις κατά καιρούς δυνατότητες της κοινωνίας για υπερνίκηση των χωρικών περιορισμών (Blumenfeld-Lieberthal, 2008; Rodrigue et al., 2013; Tsiotas, 2017). Για παράδειγμα, η δομή, η γεωμετρία και γενικά η μορφή των οδικών δικτύων είναι προφανώς διαφορετική σήμερα από ότι στο παρελθόν. Οι δομικές τους διαφορές οφείλονται στην εξέλιξη των μεταφορικών μέσων, των οδοστρωμάτων, και των αντίστοιχων τεχνολογιών τους, ενώ οι διαφορές στη μορφή τους βασίζονται σε μεταβολές που συντελούνται στην κοινωνικοοικονομική σημασία των πόλεων που συνδέουν (Rodrigue et al., 2013; Polyzos et al., 2014).

Η μελέτη των επιμέρους ιστορικών, κοινωνικοοικονομικών και χωρικών (γεωγραφικών) συνθηκών που περιβάλλουν ένα δίκτυο μεταφορών συντελεί στην απόκτηση βαθύτερης γνώσης για τη δομή και τη λειτουργικότητά του και διευκολύνει τη διαδικασία μοντελοποίησής του. Από την άλλη πλευρά, δεδομένου ότι η κατασκευή και γενικά η δημιουργία των μεταφορικών υποδομών συνιστά χρονοβόρο και δαπανηρή διαδικασία, μπορεί να θεωρηθεί ότι η μορφή και η τοπολογία των δικτύων μεταφορών επιδρούν καθοριστικά στην περαιτέρω ανάπτυξη του τομέα των μεταφορών, τόσο σε εθνικό όσο και σε διαπεριφερειακό επίπεδο (Blumenfeld-Lieberthal, 2008; Rodrigue et al., 2013; Tsiotas and Polyzos, 2015a,b). Δηλαδή, σε αντίθεση με την περίπτωση των άυλων (πχ. των κοινωνικών) δικτύων (Sgroi, 2008), το δομημένο υπόβαθρο των μεταφορικών δικτύων στερείται της ευελιξίας του επαναπροσδιορισμού της γεωμετρίας και της τοπολογίας τους, ως προς τις περιβάλλουσες κοινωνικοοικονομικές δυνάμεις, με αποτέλεσμα να προσαρμόζονται με αργούς ρυθμούς στις εκάστοτε εξελίξεις (Polyzos et al., 2014; Tsiotas and Polyzos, 2015b). Η ελαστικότητα ενός δικτύου μεταφορών να ενσωματώσει τις περιβάλλουσες εξελίξεις εξαρτάται από το σύνολο των ειδικών χαρακτηριστικών του. Για παράδειγμα, η δομή των χερσαίων δικτύων είναι περισσότερο στατική από αυτή των ακτοπλοϊκών ή των αεροπορικών, τα οποία καθίσταται ευέλικτα στον επαναπροσδιορισμό των διαδρομών τους, καθώς οι μετακινήσεις δεν πραγματοποιούνται διαμέσου υποδομών, αλλά ενός φυσικού μέσου (σε αντίθεση με τα οδικά ή σιδηροδρομικά δίκτυα) και επομένως υπόκεινται σε ασθενέστερους περιορισμούς (Rodrigue et al., 2013).

Στην Ελλάδα, οι χερσαίες μεταφορές συνιστούν βασική συνιστώσα της εθνικής και περιφερειακής οικονομίας και καθοριστικό αναπτυξιακό παράγοντα (Πολύζος, 2011), γεγονός που οφείλεται σε γεωμορφολογικά και γεωπολιτικά αίτια. Από τη μία πλευρά, η γεωπολιτική θέση της χώρας είναι καθοριστική για την ανάπτυξη του εμπορίου και των συναφών δραστηριοτήτων, ενώ η πλούσια (ορεινή και θαλάσσια) γεωμορφολογία της θέτει περιορισμούς στην ανάπτυξη των χερσαίων μεταφορών, ευνοώντας την εμφάνιση εναλλακτικών, ανταγωνιστικών, τρόπων μεταφοράς (Tsiotas and Polyzos, 2015a). Στο πλαίσιο αυτό, το παρόν άρθρο μελετά το εθνικό, διαπεριφερειακό, δίκτυο μεταφορών (Greek Road Network - GRN) της Ελλάδας, με χρήση ανάλυσης σύνθετων δικτύων (CNA) (Albert and Barabasi, 2002; Barthelemy, 2011) και εμπειρικών μεθόδων (Tsiotas and Polyzos, 2015a,b), επιδιώκοντας την εξόρυξη της κοινωνικής πληροφορίας που είναι ενσωματωμένη στην τοπολογία αυτού του δικτύου και την αξιολόγηση της συνεισφοράς του στην περιφερειακή ανάπτυξη.

Το υπόλοιπο του άρθρου οργανώνεται ως εξής: στην ενότητα 2 περιγράφεται το μεθοδολογικό πλαίσιο της μελέτης, η μοντελοποίηση του δικτύου και οι χρησιμοποιούμενες μέθοδοι ανάλυσης, η ενότητα 3 παρουσιάζει τα αποτελέσματα της ανάλυσης και το σχολιασμό τους υπό το πρίσμα της επιστήμης των Δικτύων και της Περιφερειακής Επιστήμης και, στο τέλος, στην ενότητα 4 παρατίθενται τα συμπεράσματα.

## 2. Μεθοδολογικό πλαίσιο
*2.1. Μοντελοποίηση του δικτύου*
Το εθνικό οδικό δίκτυο μεταφορών της Ελλάδας (GRN) αναπαρίσταται στον *L-χώρο* (Barthelemy, 2011; Tsiotas and Polyzos, 2015a,b) ως ένας μη κατευθυνόμενος γράφος $G(V,E)$, του οποίου το σύνολο των κόμβων $V$ αντιστοιχεί σε *διασταυρώσεις* (*intersections*) οδών, ενώ το σύνολο των ακμών $E$ αντιστοιχεί στις *οδικές διαδρομές μονής διεύθυνσης* (δίχως αλλαγή πορείας). Στο πρότυπο, οι θέσεις των κόμβων αντιστοιχούν στις επακριβείς γεωγραφικές θέσεις των διασταυρώσεων (με τις ακριβείς γεωγραφικές τους συντεταγμένες), αλλά τα μήκη των ακμών αποδίδονται ως ευθύγραμμα τμήματα και όχι με τη φυσική τους (υπό κλίμακα) μορφή. Η αντιστοίχηση αυτή αποτελεί μία συνήθη πρακτική για τη μελέτη των αστικών οδικών συστημάτων (Buhl et al., 2006; Cardillo et al., 2006), αλλά συναντάται σπανίως στην περιγραφή περιπτώσεων εθνικών οδικών δικτύων, προφανώς λόγω έλλειψης διαθεσιμότητας στοιχείων για τέτοιο επίπεδο κλίμακας. Για την περίπτωση της Ελλάδας, πραγματοποιείται για πρώτη φορά και αναμένεται να αποβεί ιδιαίτερα γόνιμη, διότι υφίσταται πλούσια βιβλιογραφία (Buhl et al., 2006; Cardillo et al., 2006; Barthelemy, 2011) προς αξιοποίηση.

Τα δεδομένα που χρησιμοποιήθηκαν για την κατασκευή του GRN αφορούν το *πρωτεύον*, το *δευτερεύον* και το *τριτεύον εθνικό* μαζί με το *πρωτεύον* και *δευτερεύον επαρχιακό* οδικό δίκτυο της Ελλάδας, όπως αυτά ορίζονται στο ΠΔ.401/93, έχουν καταρτιστεί από τη Διεύθυνση Μελετών Έργων Οδοποιίας (ΔΜΕΟ) του Υπουργείου Υποδομών, Μεταφορών και Δικτύων (νυν Υπουργείου Οικονομίας, Υποδομών, Ναυτιλίας και Τουρισμού) και διατίθενται ελεύθερα σε αρχείο χωρικής μορφής shapefile ($^*$.shp) από τον Οργανισμό Κτηματολογίου και Χαρτογραφήσεων Ελλάδος (ΟΚΧΕ, 2005). Στο παραπάνω πλαίσιο, το GRN κατασκευάστηκε ως ένας μη κατευθυνόμενος γράφος $G(V,E)$, με χωρικά βάρη (spatial weights), αποτελούμενος από $n=4.993$ κόμβους (κορυφές) και $n=6.487$ ακμές (συνδέσεις).

Το GRN προέκυψε *μη συνδετικό* ή *ασύνδετο* (*disconnected*) δίκτυο (Koschutzki et al., 2005; Tsiotas and Polyzos, 2015a), έχοντας ως συνιστώσες τα υποδίκτυα των νησιωτικών συμπλεγμάτων της χώρας, μαζί με κάποιες περιπτώσεις απομονωμένων οδικών τμημάτων που βρίσκονται στην ηπειρωτική χώρα. Το πρωτογενές αρχείο ($^*$.shp) του ΔΜΕΟ, πριν τη μετατροπή του σε γράφο, υποβλήθηκε σε ενοποίηση του συνόλου των ακμών (εντολή: merge) και μετέπειτα σε κατακερματισμό της ενιαίας γραμμικής οντότητας στα ευθύγραμμα τμήματα που το απαρτίζουν (εντολή: explode), προκειμένου να εξαλειφθούν σφάλματα της ψηφιοποίησης (όπως επικαλύψεις, διπλοεγγραφές, κλπ.).

Τέλος, το GRN αποτέλεσε *μη κατευθυνόμενο* (*non directed*) γράφο (Tsiotas and Polyzos, 2013), διότι οι ακμές αντιπροσωπεύουν τμήματα διπλής κατεύθυνσης του οδικού δικτύου, προσδίδοντας απόλυτη συμμετρία στον πίνακα συνδέσεων (συμμετρικός πίνακας). Λόγω έλλειψης δεδομένων σχετικά με τον αριθμό των λωρίδων κυκλοφορίας ανά ακμή ή με την ταχύτητα μετακίνησης στις ακμές του δικτύου, το GRN κατασκευάστηκε με μοναδικά βάρη τις χιλιομετρικές αποστάσεις των οδικών τμημάτων, με αποτέλεσμα να συνιστά ένα δίκτυο με χωρικά βάρη (spatial network). Εκτός από το μήκος των ακμών που εκφράζει τις Ευκλείδειες χιλιομετρικές αποστάσεις μεταξύ των κόμβων, η πληροφορία του φυσικού μήκους των οδικών αποστάσεων του GRN καταχωρήθηκε ως τιμή βάρους σε κάθε ακμή και παρουσιάζεται στο πρότυπο με πάχος ανάλογο του μεγέθους της.

*2.2. Μέτρα ανάλυσης δικτύου*

Τα μέτρα χώρου και τοπολογίας που χρησιμοποιούνται στην ανάλυση του GRN παρουσιάζονται συνοπτικά στον πίνακα 1.

**Πίνακας 1**
Μέτρα χώρου και τοπολογίας που χρησιμοποιούνται στην ανάλυση του GRN

| Μέτρο[(*)] | Περιγραφή | Μαθηματική Έκφραση | Αναφορά |
|---|---|---|---|
| Πυκνότητα γράφου - Graph density ($\rho$) | Ο λόγος του αριθμού των υφιστάμενων συνδέσεων (ακμών) του δικτύου $|E(G)|$ προς τον αριθμό των δυνατόν συνδέσεων που μπορούν να σχηματιστούν από το σύνολο των κόμβων $|E(G_{complete})|$. Το μέγεθος της πυκνότητας αντιπροσωπεύει την πιθανότητα εμφάνισης μιας σύνδεσης μεταξύ δύο τυχαίων κόμβων στο δίκτυο. | $\rho = \dfrac{|E(G)|}{|E(G_{complete})|} =$ $= m \Big/ \binom{n}{2} = \dfrac{2m}{n\cdot(n-1)}$ | (Tsiotas and Polyzos, 2015a) |
| Βαθμός κόμβου - Node Degree ($k$) | Ο αριθμός των προσκείμενων ακμών $e_{ij}$ σε μία κορυφή $i$ του δικτύου, ο οποίος αντιπροσωπεύει τη συνδετικότητα και την ικανότητα επικοινωνίας του δικτύου. | $k_i = k(i) = \sum_{j \in V(G)} \delta_{ij}$, $\delta_{ij} = \begin{cases} 1, & e_{ij} \in E(G) \\ 0, & \text{διαφορετικά} \end{cases}$ | (Koschutzki et al., 2005) |
| Χωρική ισχύς - Node (spatial) strength ($s$) | Το άθροισμα των χωρικών αποστάσεων $d_{ij}$ των ακμών που $e_{ij}$ πρόσκεινται σε έναν κόμβο $i$. | $s_i = s(i) = \sum_{j \in V(G)} d_{ij}$, $\delta_{ij} = \begin{cases} 1, & e_{ij} \in E(G) \\ 0, & \text{otherwise} \end{cases}$ | (Barthelemy, 2011) |
| Μέσος βαθμός κόμβων - Average Network's Degree $\langle k \rangle$ | Ο μέσος όρος των τιμών του βαθμού των κόμβων ($k_i$) για το σύνολο των κορυφών $V(G)$ του δικτύου. | $\langle k \rangle = \dfrac{1}{|V(G)|} \cdot \sum_{i=1}^{|V(G)|} k(i) =$ $= \dfrac{1}{n} \cdot \sum_{i=1}^{n} k(i)$ | (Barthelemy, 2011) |
| Κεντρικότητα εγγύτητας - Closeness Centrality[(*)] ($C_i^c$) | Ισούται με το αντίστροφο μέσο μήκος των ελάχιστων μονοπατιών ($\sum_{j=1, i \neq j}^{n} d_{ij}$) που ξεκινούν από έναν δεδομένο κόμβο $i \in V(G)$ και εκφράζει την προσβασιμότητα του κόμβου αυτού προς τους υπόλοιπους κόμβους του δικτύου. | $C_i^c = \dfrac{n-1}{\sum_{j=1, i \neq j}^{n} d_{ij}} = \langle d_{ij} \rangle^{-1}$ | (Koschutzki et al., 2005; Tsiotas and Polyzos, 2013). |
| Ενδιαμέσου κεντρικότητα - | Ισούται με το λόγο του αριθμού των ελάχιστων μονοπατιών $\sigma(k)$ του | $C_k^b = \sigma(k)/\sigma$ | (Koschutzki et al., 2005) |

| Μέτρο[*] | Περιγραφή | Μαθηματική Έκφραση | Αναφορά |
|---|---|---|---|
| *Betweenness Centrality*[*] ($C_k^B$) | δικτύου, τα οποία περιλαμβάνουν μία δεδομένη κορυφή *k*, προς το συνολικό αριθμό *σ* των μονοπατιών του δικτύου. | | |
| *Συντελεστής συγκέντρωσης - Clustering Coefficient* ($C_v$) | Εκφράζει την πιθανότητα εύρεσης συνδεδεμένων γειτόνων σε έναν τυχαίο κόμβο του δικτύου, η οποία ισοδυναμεί με το λόγο του αριθμού των συνδεδεμένων γειτόνων *E(v)* της κορυφής, προς τον αριθμό των συνολικών τριπλετών που σχηματίζονται από τη συγκεκριμένη κορυφή. | $c_v = \dfrac{\tau\rho i\gamma\omega\nu\alpha(v)}{\tau\rho\iota\pi\lambda\acute{\epsilon}\tau\epsilon\varsigma(v)} = \dfrac{E(v)}{k_v \cdot (k_v - 1)}$ | (Barthelemy, 2011; Tsiotas and Polyzos, 2015a) |
| *Συναρμολογησι μότητα - Modularity* (*Q*) | Αντικειμενική συνάρτηση που εκφράζει τη δυνατότητα διαχωρισμού του δικτύου σε κοινότητες, όπου το $g_i$ αντιπροσωπεύει την κοινότητα του κόμβου $v_i$, το [$A_{ij}$ - $P_{ij}$] τη διαφορά του παρατηρούμενου μείον τον αναμενόμενο αριθμό των ακμών που προσπίπτουν σε ένα δεδομένο ζεύγος κορυφών $v_i,v_i$ του δικτύου και $δ(g_i,g_j)$ είναι η δείκτρια συνάρτηση που επιστρέφει την τιμή 1 όταν $g_i=g_j$. | $Q = \dfrac{\sum_{i,j}[A_{ij} - P_{ij}] \cdot \delta(g_i, g_j)}{2m}$ | (Blondel et al., 2008; Fortunato, 2010) |
| *Μέσο μήκος μονοπατιού - Average Path Length* $\langle l \rangle$ | Το μέσο μήκος του ελάχιστου αριθμού των ακμών $e_{ij}$ που παρεμβάλλονται για τη σύνδεση δύο τυχαίων κορυφών του δικτύου. | $\langle l \rangle = \dfrac{\sum_{v \in V(G)} d(v_i, v_j)}{n \cdot (n-1)}$ | (Barthelemy, 2011) |

\* Όταν το μέγεθος υπολογίζεται σε δυαδικές (τοπολογικές) αποστάσεις θεωρείται *δυαδικό* (*binary* measure) και συμβολίζεται με το δείκτη *bin*, ενώ όταν υπολογίζεται σε χωρικές αποστάσεις (μετρούμενες σε ναυτικά μίλια) θεωρείται χωρικά σταθμισμένο (*weighted* measure) και συμβολίζεται με το δείκτη *wei*)

(πηγή: ίδια επεξεργασία)

Εκτός από τα βασικά προαναφερόμενα μέτρα, στην ανάλυση του GRN χρησιμοποιούνται δύο επιπλέον μεγέθη. Το πρώτο αφορά το δείκτη $r_n$ των Courtat et al. (2010), για την αξιολόγηση του επιπέδου οργάνωσης των πόλεων. Οι συγγραφείς, κατά την έρευνά τους πάνω σε αστικά συστήματα, παρατήρησαν ότι σε πολλές περιπτώσεις η κατανομή βαθμού τείνει να παρουσιάζει όξυνση στην περιοχή τιμών 3-4. Από την παρατήρηση αυτή, εισήγαγαν το δείκτη οργάνωσης των πόλεων, ο οποίος ορίζεται από τη σχέση:

$$r_n = \frac{n(1) + n(3)}{\sum_{k \neq 2} n(k)} \tag{1}$$

όπου το *n(k)* εκφράζει τον αριθμό των κόμβων που έχουν βαθμό ίσο με *k*.

Όταν ο δείκτης είναι μικρός ($r_n \approx 0$), τότε αριθμός των αδιεξόδων και των ατελών διασταυρώσεων υπολείπεται του αριθμού των κανονικών διασταυρώσεων και το αστικό σύστημα αντιστοιχεί σε ένα καλά οργανωμένο πρότυπο. Στην αντίθετη περίπτωση που ο δείκτης είναι μεγάλος και πλησιάζει τη μονάδα ($r_n \approx 1$), τότε ο αριθμός των αδιεξόδων και των ατελών διασταυρώσεων υπερτερεί του αριθμού των κανονικών διασταυρώσεων και το το αστικό σύστημα χαρακτηρίζεται από έλλειψη σχεδιασμού (Courtat et al., 2010; Barthelemy, 2011).

Το δεύτερο μέτρο αποτελεί των *ωμέγα (ω) δείκτη* των Telesford et al. (2011), το οποίο χρησιμοποιείται για την ανίχνευση της ιδιότητας του μικρού-κόσμου (small-world) *S-W* και επαγωγικά την ύπαρξη χαρακτηριστικών δικτυώματος (lattice-like characteristics) και τυχαίου γράφου (random-like characteristics). Το μέτρο αυτό συγκρίνει τη μέση συγκέντρωση του εξεταζόμενου δικτύου $\langle c \rangle$ με αυτήν ενός ισοδύναμου δικτυώματος $\langle c \rangle_{latt}$ και το μέσο μήκος μονοπατιού $\langle l \rangle$ του δικτύου με το αντίστοιχο μέγεθος ενός ισοδύναμου τυχαίου γράφου $\langle l \rangle_{rand}$, σύμφωνα με τη σχέση:

$$\omega = \left(\frac{\langle l \rangle_{rand}}{\langle l \rangle}\right) - \left(\frac{\langle c \rangle}{\langle c \rangle_{latt}}\right) \quad (2)$$

Οι τιμές του ω δείκτη που βρίσκονται κοντά στο μηδέν περιγράφουν την ιδιότητα του μικρού-κόσμου, ενώ οι θετικές τιμές υποδηλώνουν την ύπαρξη τυχαίων χαρακτηριστικών στο δίκτυο και οι αρνητικές την ύπαρξη χαρακτηριστικών δικτυώματος (Tsiotas and Polyzos, 2015β). Τα μηδενικά πρότυπα (null models) που χρησιμοποιούνται για τον υπολογισμό της παραπάνω σχέσης δημιουργούνται με χρήση των αλγορίθμων παραγωγής τυχαίων γράφων, των Maslov and Sneppen (2002), και δικτυώματος, των Sporns and Kotter (2004), οι οποίοι είναι *επαναληπτικοί* (*iterative*) και διατηρούν την κατανομή βαθμού του πρότυπου (εμπειρικού) δικτύου. Ο πρώτος εφαρμόζεται σε δύο βήματα, αρχικά επιλέγονται τυχαία τέσσερις κόμβοι των οποίων οι ακμές διχοτομούνται, αντιστοιχίζοντας μισή ακμή σε κάθε κόμβο, και στη συνέχεια οι μισές ακμές ενώνονται με τυχαίο τρόπο μεταξύ τους (Rubinov and Sporns, 2010). Ο αλγόριθμος παραγωγής του ισοδύναμου δικτυώματος των Sporns and Kotter (2004) (*latticization algorithm*) εφαρμόζει την ίδια διαδικασία, θέτοντας τον περιορισμό ότι η εναλλαγή των μισών ακμών πραγματοποιείται μόνο όταν ο προκύπτων πίνακας συνδέσεων έχει τις μη μηδενικές του καταχωρήσεις εγγύτερα στην κύρια διαγώνιο σε σύγκριση με την αρχική του κατάσταση (Sporns and Kotter, 2004; Rubinov and Sporns, 2010). Με τη συνθήκη αυτή προσεγγίζεται η τοπολογία δικτυώματος, καθόσον στα δικτυώματα είναι απίθανο να πραγματοποιηθούν συνδέσεις απομακρυσμένων κορυφών (Sporns and Kotter (2004; Rubinov and Sporns, 2010).

Γενικά, η *S-W* ιδιότητα εξετάζεται με μαθηματική αυστηρότητα σε μια διαθέσιμη οικογένεια γράφων, όταν ανιχνευθεί πως το $\langle l \rangle$ δεν αυξάνεται γρηγορότερα από λογαριθμικά καθώς ο αριθμός των κόμβων τείνει στο άπειρο, όταν δηλαδή $\langle l \rangle_{bin} = \mathcal{O}(\log n)$ καθώς $n \to \infty$ (Porter, 2012). Παρόλα αυτά, επιλέγεται η χρήση του προσεγγιστικού ελέγχου της ιδιότητας του μικρού-κόσμου, μέσα από τον υπολογισμό του ω δείκτη, διότι θεωρείται από το συγγράφοντα ότι πλεονεκτεί στα εξής σημεία: Το πρώτο αφορά τη διαθεσιμότητα της πληροφορίας, δεδομένου ότι δεν καθίσταται πάντοτε εφικτή η συλλογή μιας οικογένειας διαφορετικών διαχρονικών εκδοχών του ίδιου δικτύου για τον έλεγχο της *S-W* ιδιότητας με την εφαρμογή του ορισμού (Tsiotas and Polyzos, 2015a). Το δεύτερο σημείο αφορά το γεγονός ότι με τον υπολογισμό του ω δείκτη εκτός από τον έλεγχο της καθεαυτό *S-W* ιδιότητας παρέχονται περαιτέρω ενδείξεις για το αν η τοπολογία του εξεταζόμενου δικτύου διέπεται από χαρακτηριστικά τυχαίου δικτύου (random network) ή δικτυώματος (lattice network).

*2.3. Εμπειρική ανάλυση*
Στο επόμενο στάδιο της μελέτης του GRN πραγματοποιείται εμπειρική ανάλυση για την ανίχνευση συσχετίσεων μεταξύ των τοπολογικών χαρακτηριστικών του δικτύου και άλλων γνωστών κοινωνικοοικονομικών χαρακτηριστικών που το περιγράφουν. Η ανάλυση πραγματοποιείται σε ένα σύνολο μεταβλητών υποδομής του οδικού δικτύου (χωρικών, οικονομικών, δημογραφικών και μεταβλητών τουρισμού), οι οποίες έχουν κλίμακα αναφοράς το νομό. Η αναγωγή των μεταβλητών σε επίπεδο νομού, η οποία πραγματοποιήθηκε για την εφαρμογή της ανάλυσης συσχετίσεων, κατέστη αναγκαία επειδή στη φυσική κλίμακα αναφοράς (δηλαδή ως σημεία διασταυρώσεων) οι κόμβοι του GRN δεν έχουν κάποια περαιτέρω αξιοποιήσιμη φυσική ή οικονομική σημασία. Οι μεταβλητές που λαμβάνουν μέρος στην ανάλυση συσχετίσεων παρουσιάζονται, ανά κατηγορία, στον πίνακα 3.

**Πίνακας 2**
Μεταβλητές[*] που λαμβάνουν μέρος στην ανάλυση συσχετίσεων του GRN

| Α/Α | ΣΥΜΒΟΛ. | ΠΕΡΙΓΡΑΦΗ | ΠΗΓΗ |
|---|---|---|---|
| \multicolumn{4}{l}{Μεταβλητές Υποδομής Δικτύου} | | | |
| | LENGTH | Το συνολικό μήκος του οδικού δικτύου που περιλαμβάνεται σε κάθε νομό. | ΟΚΧΕ (2005); ίδια επεξεργασία |
| | AREA | Επιφάνεια του νομού (σε $m^2$). | ΟΚΧΕ (2005); ίδια επεξεργασία |
| | DENSITY | *Πυκνότητα δικτύου*: Πυκνότητα οδικού δικτύου ανά νομό. Ορίζεται ως το συνολικό μήκος του οδικού δικτύου που περιλαμβάνεται σε κάθε νομό προς την επιφάνεια του νομού. | ΟΚΧΕ (2005); ίδια επεξεργασία |
| | PORTS | *Αριθμός λιμένων*: Αριθμός λιμένων που περιλαμβάνονται στο νομό. | Tsiotas and Polyzos (2015a) |
| Χωρικο-οικονομικές μεταβλητές | | | |
| | IPP | *Έμμεσο πληθυσμιακό δυναμικό*: Μέτρο του όγκου των οικονομικών δραστηριοτήτων στις οποίες μία περιφέρεια έχει δυνατότητα προσέγγισης. | Πολύζος (2011) |
| | DPP | *Άμεσο ή ίδιο πληθυσμιακό δυναμικό*: Μέτρο του όγκου των οικονομικών δραστηριοτήτων που αναπτύσσονται εντός μιας περιφέρειας. | Πολύζος (2011) |
| Οικονομικές μεταβλητές | | | |
| | GDP | *Ακαθάριστο εγχώριο προϊόν νομού*: Ποσοστιαία συνεισφορά του νομού στο Ακαθάριστο Εθνικό Προϊόν. | Πολύζος (2011) |
| | $A_{SEC}$ | Συμμετοχή του *προϊόντος του πρωτογενή τομέα* του νομού στη διαμόρφωση του ΑΕΠ της χώρας, για το έτος 2013. | Tsiotas and Polyzos (2015a) |
| | $C_{SEC}$ | Συμμετοχή του *προϊόντος του τριτογενή τομέα* του νομού στη διαμόρφωση του ΑΕΠ της χώρας, για το έτος 2013. | Tsiotas and Polyzos (2015a) |
| | $AGR_{INV}$ | *Επενδύσεις αγροβιομηχανίας*: Κατά κεφαλήν ποσό που επενδύθηκε για τη δημιουργία νέων επιχειρήσεων αγροβιομηχανίας την περίοδο 2004-2008. | Polyzos et al. (2015) |
| | RPD | *Παραγωγικός δυναμισμός περιφέρειας*: σύνθετος δείκτης που συνυπολογίζει ποσά απασχόλησης, επίπεδο και δομές παραγωγής στην τοπική οικονομία. | Polyzos et al. (2015) |
| Κοινωνικές/ Δημογραφικές μεταβλητές | | | |
| | POP | *Πληθυσμός* του νομού (απογραφή 2011). | Tsiotas and Polyzos (2015a) |
| | WELF | Δείκτης *ευημερίας* του κάθε νομού. | Πολύζος (2011) |
| | EDU | Δείκτης *επιπέδου εκπαίδευσης* του πληθυσμού του νομού. | Πολύζος (2011) |
| | URB | *Δείκτης αστικοποίησης νομού*: αντιστοιχεί στον πληθυσμό της πρωτεύουσας κάθε νομού. | Polyzos et al. (2015) |
| Μεταβλητές Τουρισμού | | | |
| | $T_{GDP}$ | Συμμετοχή του *προϊόντος του τουρισμού* ενός νομού στη διαμόρφωση του ΑΕΠ της χώρας, για το έτος 2013. | Tsiotas and Polyzos (2015a) |
| | R | *Συντελεστής μεγέθυνσης* κύκλου ζωής τουριστικών περιοχών (TALC), ο οποίος εκφράζει το επίπεδο κορεσμού του νομού σε *διανυκτερεύσεις ανά* | Polyzos et al. (2013) |

| Α/Α | ΣΥΜΒΟΛ. | ΠΕΡΙΓΡΑΦΗ | ΠΗΓΗ |
|---|---|---|---|
| | | επισκέπτη. | |
| | $R_T$ | *Συντελεστής μεγέθυνσης* TALC, ο οποίος εκφράζει το επίπεδο κορεσμού του νομού σε *αριθμό επισκεπτών*. | Polyzos et al. (2013) |
| | $R_{ST}$ | *Συντελεστής μεγέθυνσης* TALC, ο οποίος εκφράζει το επίπεδο κορεσμού του νομού σε *αριθμό διανυκτερεύσεων*. | Polyzos et al. (2013) |

[*]. Μεταβλητές με στοιχεία τις τιμές του κάθε νομού για την αναγραφόμενη ιδιότητα/ χαρακτηριστικό

Στην ανάλυση συσχετίσεων υπολογίστηκε ο διμεταβλητός συντελεστής συσχέτισης του *Pearson* (*Pearson's bivariate coefficient of correlation*) $r_{xy}$ (Norusis, 2004), για κάθε ζεύγος μεταβλητών **x,y** του πίνακα 2, ο οποίος ορίζεται σύμφωνα με τη σχέση:

$$r_{xy} = \frac{Cov(\mathbf{x},\mathbf{y})}{\sqrt{Var(\mathbf{x})} \cdot \sqrt{Var(\mathbf{y})}} = \frac{Cov(\mathbf{x},\mathbf{y})}{\sigma_{\mathbf{x}} \cdot \sigma_{\mathbf{y}}} \qquad (2)$$

όπου *Cov*(**x,y**) είναι η συνδιακύμανση και *Var*(**x**)=$\sigma_{\mathbf{x}}$, *Var*(**y**)=$\sigma_{\mathbf{y}}$ οι διασπορές των διανυσματικών μεταβλητών **x,y** αντίστοιχα.

Λαμβάνοντας υπόψη ότι τα μεγέθη των νομών της Αττικής και της Θεσσαλονίκης διαφοροποιούνται πολλές φορές από την κλίμακα των υπολοίπων, παρουσιάζοντας ακραία συμπεριφορά (outlier values) (Tsiotas and Polyzos, 2013), η ανάλυση συσχετίσεων πραγματοποιείται σε δύο άξονες. Ο πρώτος περιλαμβάνει τις τιμές των μητροπολιτικών νομών και οι υπολογισμοί πραγματοποιούνται σε μεταβλητές διάστασης $n_1$=51, ενώ ο δεύτερος εξαιρεί τις εν λόγω τιμές, πραγματοποιώντας τους υπολογισμούς σε μεταβλητές διάστασης $n_2$=49.

## 3. Αποτελέσματα και συζήτηση
*3.1. Υπολογισμός των μέτρων δικτύου (network measures)*
Τα αποτελέσματα υπολογισμού των μέτρων δικτύου για το GRN παρουσιάζονται συγκεντρωτικά στον πίνακα 3. Όπως προκύπτει, το GRN είναι ένα δίκτυο με 156 συνιστώσες (components), χωρίς απομονωμένους κόμβους (isolated nodes) ($k_{GRN,min} \neq 0$) και χωρίς την ύπαρξη αυτοσυνδέσεων (self-connections), δηλαδή ακμών που συνδέουν τον ίδιο κόμβο $\left(n(e_{ii} \in E) = 0\right)$. Ο μέγιστος βαθμός του δικτύου είναι $k_{GRN,max}$=8 και είναι σχεδόν ο μισός σε σύγκριση με τις περιπτώσεις των αστικών συστημάτων (Buhl et al., 2006; Barthelemy, 2011), όπου ισχύει $k_{GRN,max} \approx 20$.

**Πίνακας 3**
Αποτελέσματα υπολογισμού των μέτρων δικτύου για το GRN

| Μετρική/ Μέγεθος | Σύμβολο | Μονάδα | Τιμή |
|---|---|---|---|
| Αριθμός κόμβων | $n$ | #[(α)] | 4.993 |
| Αριθμός ακμών | $m$ | # | 6.487 |
| Κόμβοι με αυτοσυνδέσεις | $n(e_{ii} \in E)$ | # | 0 |
| Πλήθος απομονωμένων κόμβων | $n(k=0)$ | # | 0 |
| Συνδετικές συνιστώσες | $α$ | # | 156 |
| Μέγιστος βαθμός κόμβων | $k_{max}$ | # | 8 |
| Ελάχιστος βαθμός κόμβων | $k_{min}$ | # | 1 |
| Μέσος βαθμός κόμβων | $\langle k \rangle$ | # | 2,598 |
| Μέσος σταθμισμένος βαθμός κόμβων | $\langle k_w \rangle$ | km | 14,108 |
| Μέσο μήκος ακμών | $\langle d(e_{ij}) \rangle$ | km | 5,388 |
| Συνολικό μήκος ακμών | $\sum_{ij} d(e_{ij})$ | km | 35.860 |

| Μετρική/ Μέγεθος | Σύμβολο | Μονάδα | Τιμή |
|---|---|---|---|
| Μέσο μήκος μονοπατιού | $\langle l \rangle$ | # | 46,794 |
| Μέσο μήκος μονοπατιού | $d(\langle l \rangle)$ | km | 247,52 |
| Διάμετρος δικτύου (δυαδική) | $d_{bin}(G)$ | # | 144 |
| Διάμετρος δικτύου (χιλιομετρική) | $d_w(G)$ | km | 993 |
| Πυκνότητα γράφου (επίπεδου) | $\rho_1$ | net[β] | 0,433 |
| Πυκνότητα γράφου (μη επίπεδου) | $\rho_2$ | net | 0,001 |
| Συντελεστής συγκέντρωσης | $C$ | net | 0,042 |
| Μέσος συντελεστής συγκέντρωσης | $\langle C \rangle$ | net | 0,114 |
| Συναρμολογησιμότητα | $Q$ | net | 0,946 |

α. Πλήθος στοιχείων
β. Αδιάστατος αριθμός
(πηγή: ίδια επεξεργασία)

Αντίθετα, η μέση τιμή του βαθμού $\langle k \rangle_{GRN}$=2,598 φαίνεται πως συμφωνεί με τη γενική περίπτωση $\langle k \rangle \approx 2,5$ της μελέτης των αστικών συστημάτων (Buhl et al., 2006; Barthelemy, 2011). Επίσης, ο μέσος σταθμισμένος βαθμός κόμβων (average weighted degree) ισούται με $\langle k_w \rangle_{GRN}$=14,108km και εκφράζει το συνολικό μήκος των συνδέσεων που έχει ένας τυχαίος κόμβος του δικτύου. Το μέσο μήκος ακμών (average edge length) $\langle d(e_{ij}) \rangle$ του GRN εκφράζει ότι η μέση πορεία του εθνικού (διαπεριφερειακού) οδικού δικτύου μέχρι την εύρεση της πρώτης διασταύρωσης ισούται με 5,338km, τιμή η οποία διαισθητικά φαίνεται ότι είναι αρκετά μικρή για δίκτυο εθνικής κλίμακας και προφανώς επηρεάζεται από το ηπειρωτικό ανάγλυφο της χώρας. Το συνολικό μήκος του εθνικού οδικού δικτύου της Ελλάδας, όπως προκύπτει από την άθροιση των ακμών του GRN ισούται με 35.860km. Αντιστοίχως, οι τιμές που αναφέρονται στο μέγεθος του μέσου μήκους μονοπατιού (average path length) του GRN εκφράζουν ότι η διαδρομή που παρεμβάλλεται μεταξύ δύο τυχαίων κόμβων του δικτύου αποτελείται από $\langle l \rangle$=46,794 ακμές και έχει μήκος $d(\langle l \rangle)$=247,52km.

Οι δύο παραπάνω εκδοχές του $\langle l \rangle$ αντιπροσωπεύουν το γενικευμένο κόστος των μετακινήσεων που συντελούνται εντός του δικτύου (Tsiotas and Polyzos, 2015a,b), οι οποίες φαίνεται πως επηρεάζονται από τους περιορισμούς της επιπεδότητας του δικτύου. Παρόλα αυτά, η δυαδική τιμή $\langle l \rangle_{GRN}$ είναι σαφώς μικρότερη από την αντίστοιχη τιμή ενός ισοκομβικού δικτυώματος $\langle l \rangle_{latt} = \sqrt{n} = \sqrt{4993} \approx 70,66$, γεγονός που υποδηλώνει ότι η τοπολογία του GRN εμφανίζεται περισσότερο αποτελεσματική (ως προς το πλήθος των διαδοχικών ακμών ενός μονοπατιού) από την αντίστοιχη ενός δισδιάστατου δικτυώματος.

Έπειτα, η δυαδική διάμετρος του GRN ισούται με 144 βήματα διαχωρισμού (ακμές) και εκφράζει ότι η πιο απομακρυσμένη τοπολογική (δυαδική) απόσταση του δικτύου συντίθεται από 144 ακμές. Επιπρόσθετα, η χωρική (χιλιομετρική) διάμετρος του δικτύου ισούται με 993km και εκφράζει τη χιλιομετρική απόσταση μεταξύ των δύο περισσότερο απομακρυσμένων κόμβων στο δίκτυο. Περαιτέρω, αν το GRN θεωρηθεί ως επίπεδος γράφος, τότε η πυκνότητά του υπολογίζεται από τη σχέση $m/(3n-6)$ (Barthelemy, 2011) και ισούται με $\rho_1$=0,433. Η τιμή αυτή είναι ιδιαίτερα ικανοποιητική, καθόσον περιγράφει ότι το δίκτυο περιέχει το 43,3% των δυνατών συνδέσεων που θα μπορούσαν να αναπτυχθούν στο επίπεδο για το δεδομένο αριθμό κόμβων $n_{GRN}$. Από την άλλη πλευρά, αν το GRN θεωρηθεί ως μη επίπεδος γράφος, τότε η πυκνότητά του υπολογίζεται σε $ρ_2$=0,001, τιμή η οποία είναι απειροστά μικρή και δεν φαίνεται να επιδέχεται περαιτέρω ερμηνείας, δεδομένης της επίπεδης φύσης του δικτύου.

Η τιμή του συντελεστή συγκέντρωσης (clustering coefficient) του ελληνικού εθνικού οδικού δικτύου ισούται με $C_{GRN}$=0,042, ενώ η τιμή του μέσου συντελεστή συγκέντρωσης των κόμβων είναι $\langle C \rangle_{GRN}$ =0,114. Προφανώς, η τιμή $\langle C \rangle_{GRN}$ είναι κατά πολύ μεγαλύτερη από την αντίστοιχη τιμή ενός τυχαίου δικτύου ER, η οποία προσεγγίζεται από τη σχέση $\langle C \rangle_{ER}$ ~ $1/n$=2·10$^{-4}$ (Barthelemy, 2011), γεγονός το οποίο υποδηλώνει ότι το GRN απέχει πάρα πολύ από το να περιγράφεται από την τυπολογία του τυχαίου προτύπου.

Τέλος, η τιμή της συναρμολογησιμότητας (modularity) του GRN ισούται με $Q_{GRN}$=0,946 και εκφράζει την ικανότητα επιμερισμού του δικτύου σε κοινότητες. Ως μέγεθος δικτύου (global network measure), το μέτρο αυτό παρέχει συνήθως πληροφορία όταν χρησιμοποιηθεί συγκριτικά, περιγράφοντας καλύτερη ικανότητα επιμερισμού στις περιπτώσεις που εμφανίζουν μεγαλύτερη τιμή. Στο πλαίσιο αυτό, η τιμή της $Q_{GRN}$ δεν καθίσταται περαιτέρω αξιοποιήσιμη, αλλά παρατίθεται ως δεδομένο για μελλοντική χρήση.

*3.2. Μελέτη της τοπολογίας του δικτύου*

Η μελέτη της τοπολογίας του GRN ξεκινά με την εξέταση της κατανομής του βαθμού (degree distribution) των κόμβων του δικτύου. Στο σχήμα 1 παρουσιάζονται σε μετρική και λογαριθμική κλίμακα τα διαγράμματα διασποράς (scatter plots) ($k$, $n(k)$), του βαθμού των κόμβων $k$ ως προς τη συχνότητα εμφάνισης των τιμών τους $n(k)$, τα οποία περιγράφουν την κατανομή βαθμού του GRN.

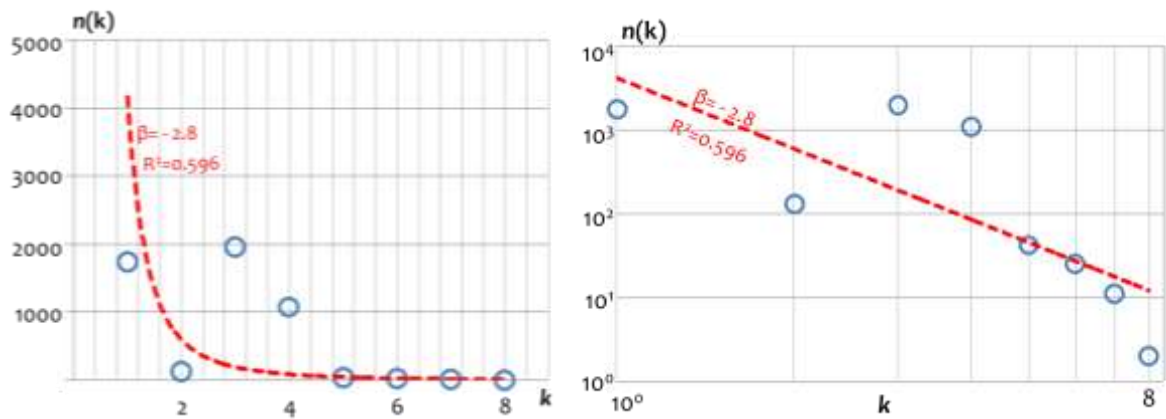

**Σχήμα 1.** Η κατανομή βαθμού ($k$, $n(k)$) του GRN, αριστερά σε μετρική και δεξιά σε λογαριθμική κλίμακα.

Όπως φαίνεται στο σχήμα 1, η κατανομή βαθμού του GRN απέχει σημαντικά από το να περιγράφει ένα πρότυπο κανόνα-δύναμης (power-law), καθόσον, με βάση την τιμή του συντελεστή προσδιορισμού ($R^2$=0,569), η διασπορά των τιμών της προσαρμόζεται σε ποσοστό 59,6% σε μία power-law καμπύλη με συντελεστή $β$=2,8. Η παρατήρηση αυτή υποδηλώνει ότι η τοπολογία του GRN δεν περιγράφεται από την ιδιότητα ελευθέρου-κλίμακας (scale-free), γεγονός το οποίο είναι αναμενόμενο και οφείλεται στην επιπεδότητα, η οποία επιβάλλει αυστηρούς περιορισμούς στο μέγεθος του βαθμού, με αποτέλεσμα η αντίστοιχη κατανομή του να παρουσιάζει όξυνση (peaked-distribution) σε μία περιοχή γύρω από τη μέση τιμή $\langle k \rangle$ (Barthelemy, 2011). Η περίπτωση του GRN είναι σύμφωνη με τα αναγραφόμενα στη σχετική βιβλιογραφία, διότι, αφενός, η τιμή του μέσου βαθμού $\langle k \rangle_{GRN}$=2,6 είναι σχεδόν ίση με την τιμή $\langle k \rangle_{GRN}$=2,5 που υπολόγισαν οι Buhl et al. (2006) στη μελέτη της τοπολογίας διαφόρων οδικών δικτύων αστικών συστημάτων παγκοσμίως και, αφετέρου, διότι η κατανομή του βαθμού *p(k)* παρουσιάζει όξυνση σε ένα εύρος τιμών 1≤*k*≤4, περιοχή που συμφωνεί με το την εμπειρική έρευνα (Barthelemy, 2011).

Για το GRN, ο δείκτης οργάνωσης των πόλεων ισούται με $r_{n,GRN}$=0,7619, τιμή η οποία περιγράφει ένα πρότυπο με άσχημη οργάνωση και έλλειψη σχεδιασμού. Βέβαια, οφείλεται να ληφθεί υπόψη ότι το GRN αποτελεί ένα διαπεριφερειακό και όχι αστικό σύστημα οδικών συνδέσεων, με αποτέλεσμα ίσως να μην καθίσταται δυνατή η τέλεια οργάνωσή του σε ένα ρυμοτομικό πλαίσιο, για την επίτευξη μικρών τιμών του εν λόγω δείκτη ($r_n$≈0). Παρόλα αυτά, η τιμή $r_{n,GRN}$ είναι ιδιαίτερα μεγάλη, ώστε να εξάγεται με σχετική ασφάλεια το συμπέρασμα της έλλειψης σχεδιασμού.

Στο επόμενο στάδιο, μελετάται η τοπολογία του GRN, μέσα από τη φασματική πληροφορία που περιέχεται στον πίνακα συνδέσεων του οδικού δικτύου. Βασικό εργαλείο στην ανάλυση αυτή αποτελούν τα διαγράμματα πυκνότητας ή σποραδικότητας (sparsity pattern plots or spy plots) (Χατζίκος, 2007), τα οποία εικονίζουν με κουκκίδες (σημεία) τις θέσεις των μη μηδενικών στοιχείων ενός πίνακα. Με τα διαγράμματα αυτά οπτικοποιείται η εσωτερική δομή ενός πίνακα και στην περίπτωση που αυτός αποτελεί πίνακα συνδέσεων παρέχεται μία σχηματική εικόνα για το πρότυπο συνδετικότητας του γράφου, εντός του δισδιάστατου χώρου που ορίζεται από την εν λόγω μητρωική διάταξη. Για την εξαγωγή συμπερασμάτων, η χρήση των διαγραμμάτων σποραδικότητας πραγματοποιείται συγκριτικά, δηλαδή με την αντιπαραβολή πινάκων που αντιστοιχούν είτε σε διαφορετικές καταστάσεις του ιδίου πίνακα είτε σε διαφορετικούς συγκρίσιμους πίνακες.

Στο σχήμα 2 παρουσιάζονται τα διαγράμματα σποραδικότητας του πίνακα συνδέσεων (a) του οδικού δικτύου της Ελλάδας (GRN), και τεσσάρων ισοκομβικών δικτύων (*n*=σταθ), τα οποία εμφανίζουν την ίδια κατανομή βαθμού με το GRN και τις ιδιότητες (b) *ελευθέρου-κλίμακας* (scale-free network), (c) *δικτυώματος* (lattice network), (d) *μικρού-κόσμου* (small-world) και (e) *τυχαίου δικτύου* (random network) αντίστοιχα. Για την κατασκευή των μηδενικών προτύπων (null models) με τις παραπάνω ιδιότητες χρησιμοποιήθηκαν οι αλγόριθμοι που περιγράφονται από τους Maslov and Sneppen (2002) και τους Sporns and Kotter (2004). Όπως φαίνεται, το πρότυπο σποραδικότητας του πίνακα συνδέσεων του GRN εμφανίζει ιδιαίτερη συγκέντρωση και συμπάγεια στις θέσεις κοντά στις διαγώνιους, γεγονός που οφείλεται προφανώς στην ύπαρξη ισχυρών χωρικών περιορισμών που καθιστούν απαγορευτική τη σύναψη συνδέσεων μεταξύ ιδιαίτερα απομακρυσμένων θέσεων. Το πρότυπο που διαμορφώνεται από τη σποραδικότητα των τιμών του πίνακα συνδέσεων του GRN φαίνεται πως ομοιάζει περισσότερο με τα πρότυπα (c) του ισοκομβικού δικτυώματος και (d) του μικρού-κόσμου, τα οποία εμφανίζουν μία ισχυρή συγκέντρωση τιμών στην κατεύθυνση της κύριας διαγωνίου, σε αντίθεση με τις άλλες δύο περιπτώσεις, οι οποίες έχουν τις τιμές τους διασπαρμένες σε όλη την έκταση του πίνακα.

Περαιτέρω, ιδιαίτερο θεωρητικό ενδιαφέρον παρουσιάζει η μορφή των διαγραμμάτων σποραδικότητας των μηδενικών προτύπων του σχήματος 2, η οποία προκύπτει συνεπής με τη θεωρία (Gilbert, 1961; Watts and Strogatz, 1998; Barabasi and Albert, 1999; Dall and Christensen, 2002, Barthelemy, 2011). Αναλυτικότερα, σύμφωνα με το σχήμα 3(b), το διάγραμμα σποραδικότητας του προτύπου ελευθέρου-κλίμακας (b) εμφανίζει ιδιαίτερη συγκέντρωση τιμών στις δύο προσκείμενες πλευρές του πίνακα συνδέσεων, οι οποίες συνθέτουν την πάνω κορυφή της κύριας διαγωνίου. Το γεγονός αυτό υποδηλώνει την ύπαρξη μιας βασικής ομάδας πλημνών στο δίκτυο και την τάση σύνδεσης των υπολοίπων κόμβων σε αυτές, μέσω του μηχανισμού της επιλεκτικής προσάρτησης (preferential attachment). Η εικόνα που παρουσιάζει το πρότυπο δικτύωμα (c) χαρακτηρίζεται από την αναμενόμενη συγκέντρωση των τιμών γύρω από την κύρια διαγώνιο του πίνακα συνδέσεων. Η εκτροπή κάποιων τιμών από την κύρια διαγώνιο και οι μικρές συγκεντρώσεις που παρατηρούνται στις κορυφές της δευτερεύουσας διαγωνίου του πίνακα συνδέσεων ενδεχομένως να σχετίζονται με την έλλειψη επιλογής του περιορισμού της επιπεδότητας στον αλγόριθμο κατασκευής του ισοκομβικού δικτυώματος (Sporns and Kotter, 2004).

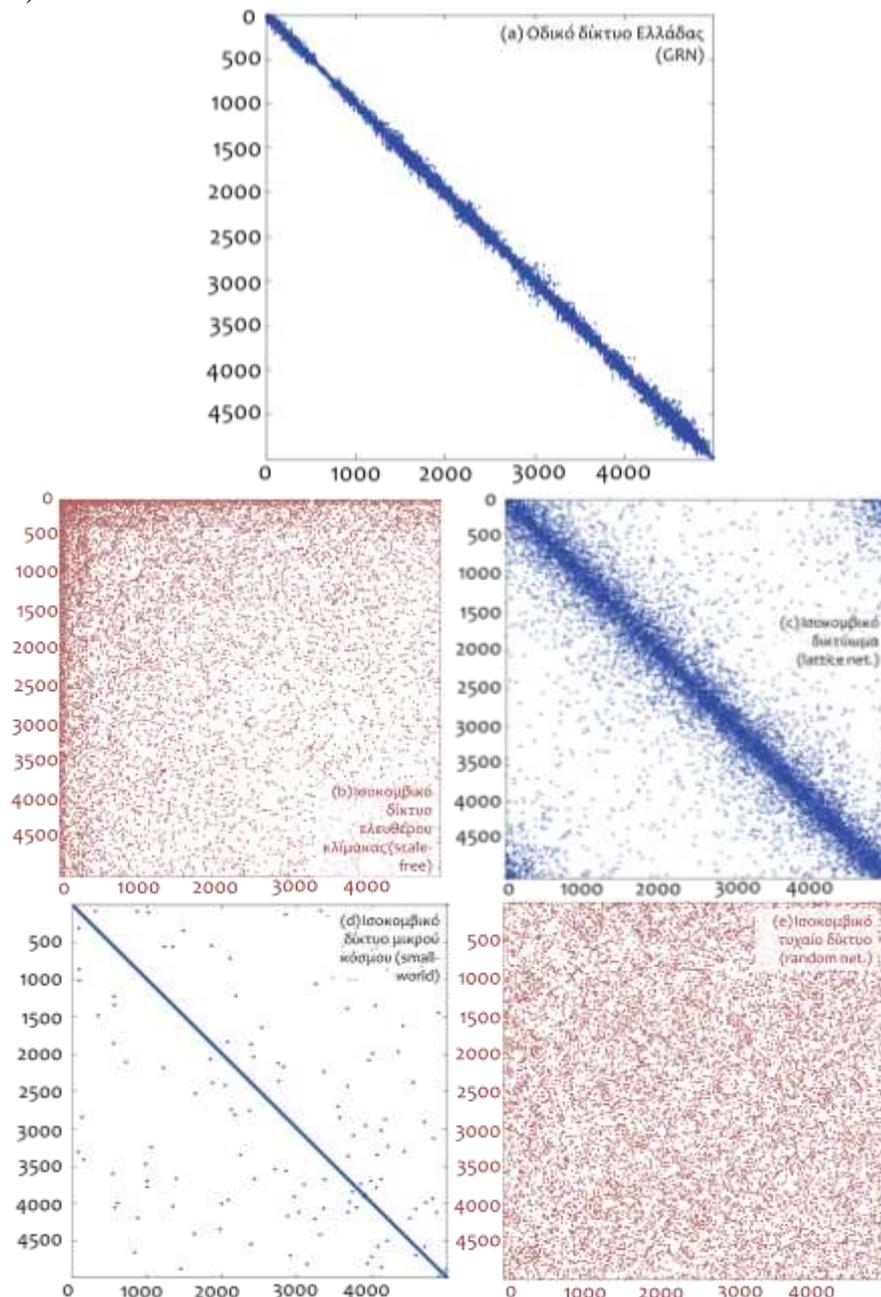

**Σχήμα 2.** Διαγράμματα σποραδικότητας (spy plots) των πινάκων συνδέσεων (adjacency matrices) (a) του οδικού δικτύου της Ελλάδας (GRN) (b) Ενός ισοκομβικού δικτύου με την ίδια κατανομή βαθμού και την ιδιότητα ελευθέρου-κλίμακας (scale-free) (c) Ενός ισοκομβικού δικτυώματος (lattice network) με την ίδια κατανομή βαθμού (d) Ενός ισοκομβικού δικτύου με την ίδια κατανομή βαθμού και την ιδιότητα του μικρού-κόσμου (small-world) και (e) Ενός ισοκομβικού τυχαίου δικτύου (random network) με την ίδια κατανομή βαθμού (οι άξονες εκφράζουν την αρίθμηση των κόμβων του δικτύου).

Τέλος, οι μορφές των διαγραμμάτων σποραδικότητας των προτύπων του μικρού-κόσμου (d) και του τυχαίου δικτύου (e) εμφανίζονται σε πλήρη συνέπεια με τη θεωρία. Από τη μια μεριά, στο πρότυπο του μικρού-κόσμου οι ελάχιστες τιμές που απέχουν από τον άξονα της κύριας διαγωνίου του πίνακα συνδέσεων υποδηλώνουν την ύπαρξη των χαρακτηριστικών συντομεύσεων (shortcuts) που προσδίδουν στο δίκτυο αυτό την καλή ικανότητα πρόσβασης σε όλα τις θέσεις του (Watts and Strogatz, 1998). Από την άλλη πλευρά, η σχεδόν ομοιόμορφη διασπορά των τιμών που παρατηρείται σε όλο το εύρος του πίνακα συνδέσεων του τυχαίου προτύπου (e) οφείλεται στη σταθερή πιθανότητα εμφάνισης των ακμών στο δίκτυο (Gilbert, 1961; Dall and Christensen, 2002, Barthelemy, 2011), η οποία συνιστά την κατασκευαστική συνθήκη του τυχαίου δικτύου.

Έπειτα, τα αποτελέσματα της προσεγγιστικής ανάλυσης για την ανίχνευση της ιδιότητας του μικρού-κόσμου στο GRN παρουσιάζονται στον πίνακα 4, σύμφωνα με τον οποίο, η τιμή του $\omega$ δείκτη προκύπτει αρνητική και ιδιαίτερα κοντά στη μονάδα ($\omega$=-0,847). Το αποτέλεσμα αυτό εκφράζει ότι το GRN διέπεται από χαρακτηριστικά δικτυώματος (lattice-like characteristics), γεγονός που είναι σύμφωνο με τη μέχρι στιγμής προηγηθείσα ανάλυση.

**Πίνακας 4**
Αποτελέσματα της προσεγγιστικής ανάλυσης
για την ανίχνευση της ιδιότητας του μικρού-κόσμου για το GRN

| Μέγεθος | $\langle c \rangle$ | $\langle c \rangle_{latt}$ | $\langle l \rangle$ | $\langle l \rangle_{rand}$ | $\omega^{*}$ |
|---|---|---|---|---|---|
| Τιμή | 0,114 | 0,108 | 46,749 | 9,737 | **-0,847** |
| Ένδειξη | Συμπεριφορά δικτυώματος (lattice-like characteristics) | | | | |

[*]. Σύμφωνα με τη σχέση (2)

Στο επόμενο στάδιο, υπολογίζονται τα βασικά μέτρα τοπολογίας και κεντρικότητας (βαθμός, ενδιαμεσότητα, εγγύτητα, συγκέντρωση, συναρμολογησιμότητα και χωρική ισχύς) των κόμβων του GRN, των οποίων η χωρική κατανομή παρουσιάζεται στους τοπολογικούς χάρτες του σχήματος 3. Αρχικά, η χωρική κατανομή του βαθμού (σχήμα 3a) εμφανίζεται αρκετά ασαφής, καθώς σε όλο το μήκος του τοπολογικού χάρτη του GRN παρατηρείται μια σχετικά ομοιόμορφη διασπορά όλων των χρωματικών διαβαθμίσεων των τιμών του βαθμού (degree). Η πιο ενδιαφέρουσα ίσως παρατήρηση αφορά το γεγονός ότι οι κεντρικές γεωγραφικές περιοχές του δικτύου (Στερεά Ελλάδα, Θεσσαλία) φαίνεται πως δεν είναι ιδιαίτερα προνομιούχες σε συνδεσιμότητα, όπως θα ήταν αναμενόμενο, προφανώς λόγω της ηπειρωτικής μορφολογίας που χαρακτηρίζει το δυτικό τμήμα της χώρας.

Αντίθετα με το βαθμό, η χωρική κατανομή των τιμών της ενδιαμέσου κεντρικότητας (betweenness centrality) $C^b$ φαίνεται πως παρέχει περισσότερο διαφωτιστικές πληροφορίες. Όπως φαίνεται στο σχήμα 3(b), οι μεγαλύτερες τιμές της κατανομής της ενδιαμέσου κεντρικότητας (με πράσινο χρώμα) συγκεντρώνονται κατά μήκος ενός άξονα που βρίσκεται στην ανατολική πλευρά της χώρας και εκτείνεται από την περιοχή της Κορίνθου μέχρι τη Θεσσαλονίκη. Είναι προφανές ότι ο άξονας των μεγάλων τιμών της ενδιαμέσου κεντρικότητας $C^b$ συμπίπτει στο μεγαλύτερο μέρος του με τον εθνικό οδικό άξονα της ΠΑΘΕ, με εξαίρεση το τμήμα της Κεντρικής Ελλάδας, στο οποίο η κατανομή των τιμών της $C^b$ είναι μετατοπισμένη προς τα δυτικά. Αυτή η αναντιστοιχία που παρατηρείται σε τοπική κλίμακα οφείλεται προφανώς στο ηπειρωτικό ανάγλυφο της χώρας, το οποίο επέβαλε την κατασκευή αυτοκινητοδρόμων στην ανατολική και σαφώς περισσότερο πεδινή πλευρά της χώρας.

Η χωρική κατανομή των τιμών της κεντρικότητας εγγύτητας (closeness centrality) $C^c$ (σχήμα 3c) εμφανίζει ένα αναμενόμενο πρότυπο. Από τη μία πλευρά, στα νησιά παρουσιάζει μεγάλες τιμές, διότι οι μέσες αποστάσεις μεταξύ κόμβων είναι σημαντικά μικρότερες σε σχέση με τις αντίστοιχες της ηπειρωτικής Ελλάδας. Από την άλλη πλευρά, η κατανομή της $C^c$ στην ηπειρωτική χώρα παρουσιάζει μικρές τιμές στις μεθόριες περιοχές (Ανατολική Μακεδονία, Θράκη, Δυτική Πελοπόννησος) και μεγάλες τιμές στον κεντρικό (ηπειρωτικό) κορμό της χώρας.

Ακολούθως, η χωρική κατανομή του συντελεστή συγκέντρωσης (clustering coefficient) $C$ (σχήμα 3d) εμφανίζεται ιδιαίτερα σύνθετη. Μεγάλες τιμές του συντελεστή φανερώνουν αλληλοσυνδεδεμένες περιοχές (ύπαρξη τριγώνων), δηλαδή περιοχές με πολλές κυκλικές συνδέσεις, στις οποίες οι κόμβοι διαθέτουν γείτονες που είναι συνδεδεμένοι μεταξύ τους. Η κατάσταση αυτή ενδεχομένως να σχετίζεται με την ύπαρξη περιοχών με σημαίνουσα οικονομική ή συναφή δραστηριότητα, καθόσον μεταξύ των θέσεων με μεγάλες τιμές υφίστανται οι περιοχές της Αχαΐας, της Αττικής, των Χανίων, της Λέσβου, των Ιωαννίνων, της Δυτικής Εύβοιας, της Λάρισας, της Ανατολικής Θεσσαλονίκης, κλπ.

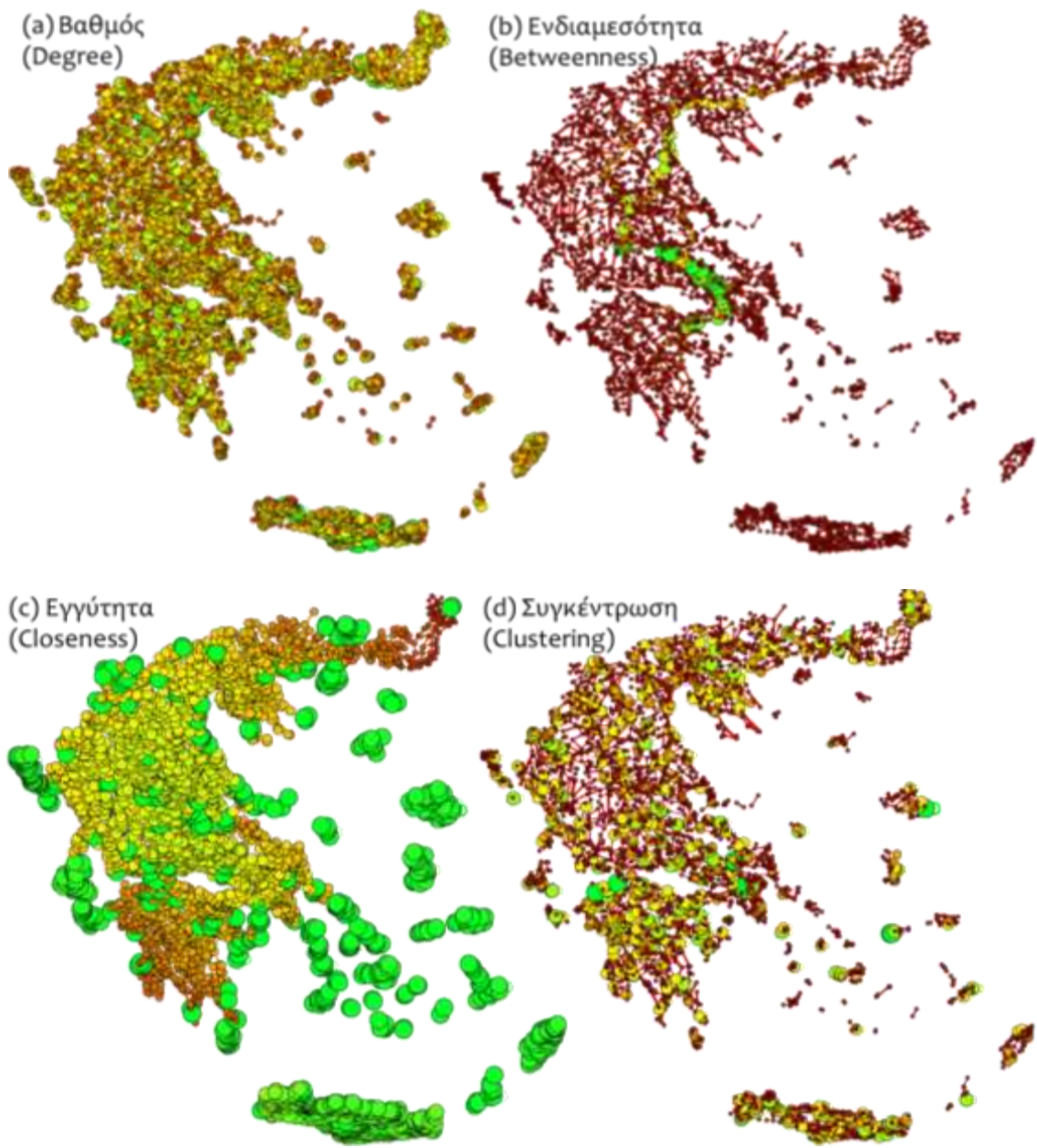

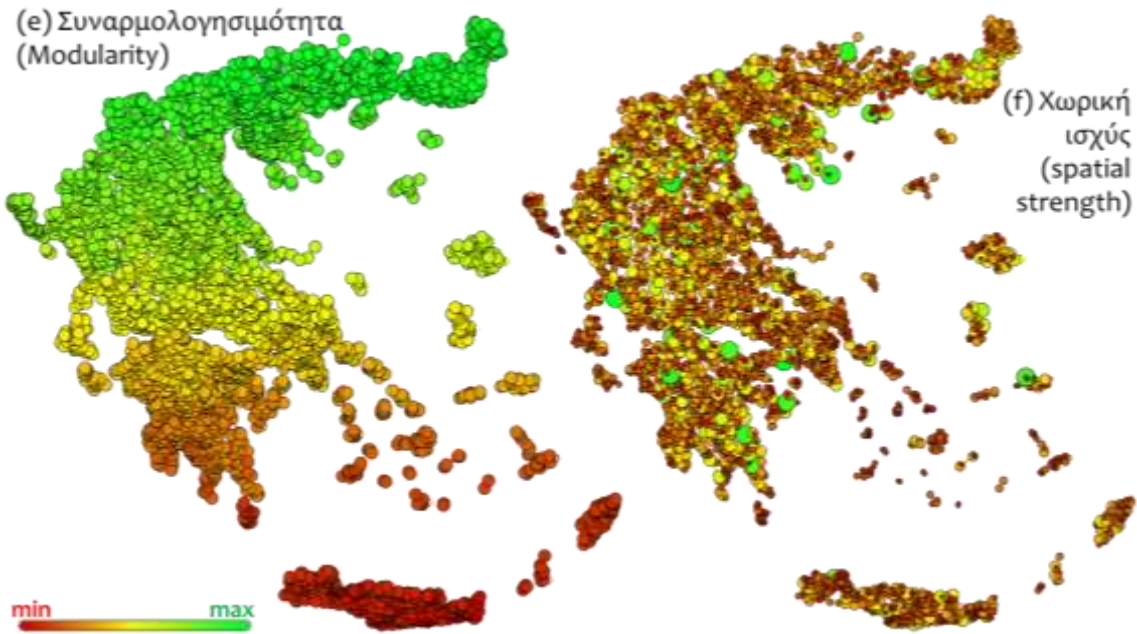

**Σχήμα 3.** Χωρική κατανομή των μέτρων κόμβου για το GRN: (a) Βαθμός (degree) (b) Ενδιαμεσότητα (betweenness) (c) Εγγύτητα (closeness) (d) Συγκέντρωση (clustering) (e) Συναρμολογησιμότητα (modularity classification) και (f) Χωρική ισχύς (spatial strength).

Έπειτα, η χωρική κατανομή των τιμών της *Q-κατηγοριοποίησης* (modularity classification) (σχήμα 3e), δηλαδή των τιμών που αντιστοιχούν στις κοινότητες που ανήκουν οι κόμβοι του δικτύου, παρουσιάζεται απόλυτα συνεπής με τη θεωρία. Η σχετική εμπειρική έρευνα έχει γενικά αναδείξει ότι ο διαμοιρασμός των χωρικών δικτύων σε κοινότητες διέπεται κατά κανόνα από γεωγραφικά κριτήρια, μη παρέχοντας ιδιαίτερα αξιοποιήσιμη δομική πληροφορία, επειδή οι σημαντικότερες ροές στο δίκτυο εντοπίζονται μεταξύ κόμβων που ανήκουν σε ίδιες ή παρόμοιες γεωγραφικές περιοχές (Guimera et al., 2005; Kaluza et al., 2010; Barthelemy, 2011). Στο παραπάνω πλαίσιο εντάσσεται και η εικόνα της χωρικής κατανομής των τιμών της *Q-κατηγοριοποίησης*, η οποία επιμερίζεται σε ευδιάκριτες οριζόντιες χρωματικές ζώνες με γεωγραφική συνάφεια, γεγονός που εμφανίζεται συνεπές με τα χαρακτηριστικά δικτυώματος που διέπουν την τοπολογία του GRN.

Τέλος, η περίπτωση της κατανομής της χωρικής ισχύος (spatial strength) *s* (σχήμα 3f) παρουσιάζει ένα σύνθετο πρότυπο. Λαμβάνοντας υπόψη ότι στο μέγεθος αυτό αθροίζονται οι χιλιομετρικές αποστάσεις των προσκείμενων σε έναν κόμβο ακμών, ως κεντρικοί (με πράσινο χρώμα) εμφανίζονται οι κόμβοι που είναι περισσότερο απομακρυσμένοι από τους γειτονικούς τους. Το γεγονός αυτό αναδεικνύει στο δίκτυο κάποιες θέσεις που, λόγω της μεγαλύτερης σχετικής απόστασης που έχουν προς τους γείτονές τους, εμφανίζουν μικρότερη εξάρτηση στην οδική τους επικοινωνία και πιθανώς μεγαλύτερη αυτονομία στις οικονομικές τους δραστηριότητες.

Στο τελευταίο στάδιο, εξετάζονται διαγραμματικά οι συσχετίσεις που παρουσιάζουν τα μεγέθη της ενδιαμέσου κεντρικότητας $C^b(k)$, της χωρικής ισχύος $s(k)$ και του συντελεστή συγκέντρωσης $C(k)$, ως προς το βαθμό *k*. Τα διαγράμματα διασποράς των δύο πρώτων περιπτώσεων $(k, C^b)$ και $(k, s)$ παρουσιάζονται στο σχήμα 4, ενώ της περίπτωσης $(k, C(k))$ παρουσιάζεται στο σχήμα 5.

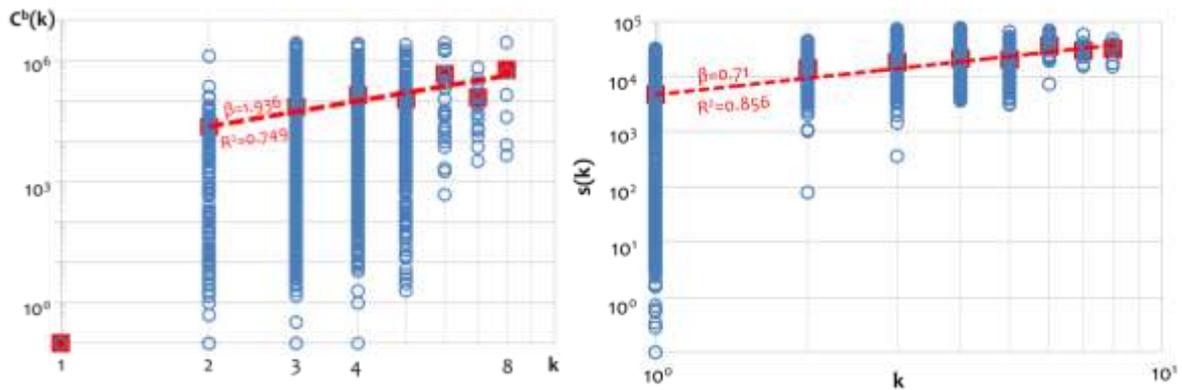

**Σχήμα 4.** Διαγράμματα διασποράς (scatter plots) (αρ.) βαθμού-ενδιαμέσου κεντρικότητας ($k$,$C(k)^b$) και (δεξ.) βαθμού-χωρικής ισχύος ($k$,$s(k)$) για το GRN. Τα κόκκινα τετράγωνα αντιστοιχούν στις μέσες τιμές για την κάθε τιμή/κατηγορία του βαθμού.

Η εικόνα που παρουσιάζει το διάγραμμα διασποράς του GRN, μεταξύ βαθμού και μέσης ενδιαμέσου κεντρικότητας ανά βαθμό ($k$, $\left\langle C^b \right|_{k=k_i} \right\rangle$), $i$=2,…,8, φαίνεται πως είναι σύμφωνη με τα αποτελέσματα της έρευνας των Crucitti et al. (2006), οι οποίοι εργαζόμενοι πάνω στην κεντρικότητα οδικών δικτύων αστικών συστημάτων έδειξαν πως η σχέση ($k$, $\left\langle C^b \right|_{k=k_i} \right\rangle$) είναι θετική και όχι απολύτως γραμμική. Στην περίπτωση του GRN η γραμμική σχέση που εικονίζεται στο σχήμα 4 αντιστοιχεί σε μία ικανοποιητική παρεμβολή ($R^2$=0,749) μιας καμπύλης κανόνα-δύναμης (power-law), η οποία έχει συντελεστή $β_{GRN}$=1,936.

Ως προς τη σχέση βαθμού και ισχύος ($k$,$s$), η εικόνα που σχηματίζεται στο GRN φαίνεται πως ομοιάζει με τα σχετικά πορίσματα της μελέτης του δικτύου commuting της Σαρδηνίας, η οποία πραγματοποιήθηκε από τους de Montis et al. (2007). Οι συγγραφείς διαπίστωσαν ότι η σχέση $\langle s \rangle$=$f(k)$ του δικτύου commuters της Σαρδηνίας περιγράφεται από το πρότυπο κανόνα-δύναμης (γραμμική σε λογαριθμική κλίμακα). Η περίπτωση του GRN συμφωνεί με αυτή την παρατήρηση, με βάση το συντελεστή προσδιορισμού $R^2$=0,856, αλλά έχει περίπου τη μισή κλίση ($β_{GRN}$=0,7) σε σχέση με την περίπτωση της Σαρδηνίας ($β_{SCN}$=1,9).

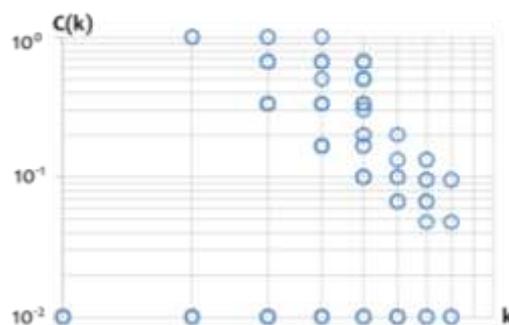

**Σχήμα 5.** Μεταβολή του συντελεστή συγκέντρωσης $C(k)$ σε σχέση με το βαθμό κόμβων $k$ του οδικού δικτύου Ελλάδας (GRN). Η μορφή του διαγράμματος διασποράς υποδεικνύει λογαριθμική μείωση όσο μεγαλώνουν οι τιμές του $k$.

Τέλος, η σχέση μεταξύ βαθμού και συντελεστή συγκέντρωσης ($k$,$C(k)$) (σχήμα 5) διέπεται από ένα μηχανισμό λογαριθμικής μείωσης με την αύξηση των τιμών του $k$, κατάσταση που παρατηρείται συχνά στα χερσαία δίκτυα (Sen et al., 2003, Barthelemy, 2011). Το γεγονός αυτό σχετίζεται προφανώς με την ύπαρξη χωρικών περιορισμών που παρέχουν το προνόμιο της συνδετικότητας σε λίγους κόμβους, με αποτέλεσμα οι περισσότεροι γείτονές τους να έχουν χαμηλή συνδεσιμότητα.

*3.3. Εμπειρική ανάλυση*

Τα αποτελέσματα της ανάλυσης παρουσιάζονται στον πίνακα 5, τα οποία διαχωρίζονται στις περιπτώσεις των υπολογισμών που πραγματοποιήθηκαν με τους μητροπολιτικούς νομούς της Αθήνας και της Θεσσαλονίκης ($n_1$=51) και χωρίς αυτούς ($n_2$=49).

**Πίνακας 5**
Αποτελέσματα της ανάλυσης συσχετίσεων για το GRN[α]

| | Με μητροπολιτικούς νομούς ($n_1$=51) | | | | Χωρίς μητροπολιτικούς νομούς ($n_2$=49) | | | |
|---|---|---|---|---|---|---|---|---|
| Μεταβλ. *y* → | LENGTH | | DENSITY | | LENGTH | | DENSITY | |
| Μεταβλητή *x*↓ | $r_{xy}$[β] | sig.[γ] | $r_{xy}$ | sig. | $r_{xy}$ | sig. | $r_{xy}$ | sig. |
| LENGTH | 1 | | 0,229 | 0,106 | 1 | | 0,207 | 0,154 |
| AREA | 0,839**[δ] | 0,000 | -0,276 | 0,050 | 0,879** | 0,000 | -0,230 | 0,112 |
| DENSITY | 0,229 | 0,106 | 1 | | 0,207 | 0,154 | 1 | |
| PORTS | 0,018 | 0,919 | 0,329 | 0,053 | -0,032 | 0,862 | -0,201 | 0,261 |
| IPP | 0,285*[ε] | 0,043 | -0,147 | 0,303 | 0,296* | 0,039 | -0,204 | 0,159 |
| DPP | 0,340* | 0,015 | 0,707** | 0,000 | 0,674** | 0,000 | 0,109 | 0,458 |
| GDP | 0,239 | 0,091 | 0,742** | 0,000 | 0,667** | 0,000 | 0,177 | 0,223 |
| $A_{SEC}$ | 0,052 | 0,718 | -0,321* | 0,022 | 0,135 | 0,354 | -0,169 | 0,246 |
| $C_{SEC}$ | -0,085 | 0,555 | 0,324* | 0,020 | -0,123 | 0,401 | 0,326* | 0,022 |
| $T_{GDP}$ | 0,229 | 0,107 | 0,768** | 0,000 | 0,301* | 0,036 | 0,395** | 0,005 |
| $AGR_{INV}$ | -0,314* | 0,025 | -0,129 | 0,368 | -0,324* | 0,023 | -0,335* | 0,019 |
| RPD | 0,122 | 0,395 | 0,064 | 0,656 | 0,080 | 0,587 | -0,117 | 0,423 |
| POP | 0,243 | 0,086 | 0,744** | 0,000 | 0,713** | 0,000 | 0,145 | 0,321 |
| WELF | -0,009 | 0,950 | 0,514** | 0,000 | -0,101 | 0,489 | 0,334* | 0,019 |
| EDU | 0,294* | 0,036 | 0,644** | 0,000 | 0,254 | 0,079 | 0,341* | 0,017 |
| URB | -0,226 | 0,111 | -0,106 | 0,460 | -0,203 | 0,161 | 0,085 | 0,560 |
| R | -0,157 | 0,272 | 0,338* | 0,015 | -0,139 | 0,341 | 0,584** | 0,000 |
| $R_T$ | 0,071 | 0,621 | -0,074 | 0,604 | 0,085 | 0,563 | -0,062 | 0,671 |
| $R_{ST}$ | 0,320* | 0,022 | 0,651** | 0,000 | 0,280 | 0,051 | 0,668** | 0,000 |

α. Υπολογισμοί στις μεταβλητές του πίνακα 3
β. διμεταβλητός συντελεστής συσχέτισης του Pearson
γ. δίπλευρη (2-tailed) σημαντικότητα
δ. Τιμές σημαντικές (**) σε επίπεδο σημαντικότητας 0.01
ε. Τιμές σημαντικές (*) σε επίπεδο σημαντικότητας 0.05

Με μια γενική ματιά, φαίνεται πως η εξαίρεση των νομών της Αττικής και Θεσσαλονίνης ($n_2$=49) οδηγεί κατά κανόνα στην παραγωγή ευνοϊκότερων αποτελεσμάτων (μεγαλύτερες τιμές των στατιστικά σημαντικών συντελεστών) για τη μεταβλητή του μήκος του οδικού δικτύου (LENGTH), ενώ σε σημαντικές ανantiστοιχίες στην περίπτωση της μεταβλητής της πυκνότητας (DENSITY). Ειδικότερα, η μεταβλητή του *μήκους του οδικού δικτύου* (**y**=LENGTH) παρουσιάζεται να είναι σημαντικά συσχετισμένη με τις παρακάτω μεταβλητές:

- *Επιφάνεια του νομού* (**x**=*AREA*): Η περίπτωση ($r_{AREA,LENGTH}^{n_1=51} = 0,839$, $r_{AREA,LENGTH}^{n_2=49} = 0,879$) είναι τετριμμένη, καθόσον οι μεγάλες γεωγραφικές επιφάνειες διαθέτουν μεγαλύτερη χωρητικότητα για την κατασκευή οδικών έργων σε μεγαλύτερη κλίμακα. Η απόκλιση από την πλήρη γραμμικότητα $r_{AREA,LENGTH} \neq 1$ αυτού του συντελεστή προφανώς οφείλεται στο ηπειρωτικό ανάγλυφο της ενδοχώρας, το οποίο έχει αποτρέψει

Page | 17

την κατασκευή έργων οδοποιίας μεγάλης κλίμακας σε νομούς με μεγάλες ορεινές επιφάνειες.

- *Πληθυσμιακά δυναμικά* (**x**=*IPP* και **x**=*DPP*): Οι τιμές του συντελεστή συσχέτισης στις περιπτώσεις αυτές εκφράζει ότι οι νομοί με μεγάλο μήκος οδικού δικτύου τείνουν να έχουν μεγαλύτερο συγκριτικό πλεονέκτημα για οικονομική ανάπτυξη, έναντι των άλλων με χαμηλές τιμές πληθυσμιακού δυναμικού (Πολύζος, 2011). Ειδικότερα, οι τιμές του συντελεστή που υπολογίζονται στο έμμεσο πληθυσμιακό δυναμικό ($r_{IPP,LENGTH}^{n_1=51} = 0,285$, $r_{IPP,LENGTH}^{n_2=49} = 0,296$) εκφράζουν τη δυνατότητα ευκολότερης πρόσβασης των νομών με μεγάλο μήκος οδικού δικτύου στις οικονομικές δραστηριότητες των υπολοίπων, ενώ αυτές που υπολογίζονται στο άμεσο πληθυσμιακό δυναμικό ($r_{DPP,LENGTH}^{n_1=51} = 0,340$, $r_{DPP,LENGTH}^{n_2=49} = 0,674$), εκφράζουν τη δυνατότητα ευκολότερης πρόσβασης των νομών με μεγάλο μήκος οδικού δικτύου στις ίδιες οικονομικές τους δραστηριότητες.

- *Επενδύσεις στην αγροτική βιομηχανία* (**x**=$AGR_{INV}$): Η τιμή του συντελεστή συσχέτισης ($r_{AGR_{INV},LENGTH}^{n_1=51} = 0,314$, $r_{AGR_{INV},LENGTH}^{n_2=49} = 0,324$) εκφράζει την ύπαρξη αρνητικής αναλογίας μεταξύ των συσχετιζόμενων μεταβλητών. Το αποτέλεσμα αυτό εκφράζει ότι οι νομοί που διαθέτουν μεγάλο μήκος οδικού δικτύου τείνουν να μην απολαμβάνουν εξίσου μεγάλες επενδύσεις στην αγροτική παραγωγή, γεγονός το οποίο φαίνεται πως αναδεικνύει μία παθογένεια στη λειτουργία και την περαιτέρω ανάπτυξη του πρωτογενή τομέα της χώρας, λαμβάνοντας υπόψη ότι ο ρόλος των οδικών υποδομών στην υποστήριξη των αγροτικών δραστηριοτήτων είναι καθοριστικός.

- *Επίπεδο εκπαίδευσης* (**x**=*EDU*): Εδώ ο συντελεστής προκύπτει σημαντικός μόνο για τη συνολική περίπτωση ($r_{EDU,LENGTH}^{n_1=51} = 0,294$), ενώ για την περίπτωση χωρίς τους μητροπολιτικούς νομούς ($r_{EDU,LENGTH}^{n_2=49} = 0,254$) το επίπεδο σημαντικότητας αμβλύνεται στο 0,1. Το θετικό πρόσημο του συντελεστή εκφράζει ότι οι νομοί με μεγάλο μήκος οδικού δικτύου τείνουν να έχουν και σχετικά υψηλό επίπεδο εκπαίδευσης του πληθυσμού τους, αλλά η σχέση αυτή δεν επιδέχεται περαιτέρω ερμηνείας και προφανώς οφείλεται στην από κοινού επίδραση κάποιας τρίτης (λανθάνουσας) μεταβλητής.

- *Συντελεστής μεγέθυνσης TALC για τον αριθμό των διανυκτερεύσεων* (**x**=$R_{ST}$): Στην περίπτωση αυτή ο συντελεστής προκύπτει σημαντικός μόνο όταν λαμβάνονται υπόψη και οι μητροπολιτικοί νομοί ($r_{R_{ST},LENGTH}^{n_1=51} = 0,320$), αλλά και η περίπτωση δίχως τους μητροπολιτικούς νομούς ($r_{R_{ST},LENGTH}^{n_2=49} = 0,280$) δεν απέχει από το να θεωρηθεί σημαντική, καθόσον ξεπερνά οριακά το επίπεδο σημαντικότητας 0,05. Η κατάσταση αυτή υποδηλώνει πως οι νομοί με μεγάλο μήκος οδικού δικτύου τείνουν να διατηρούν για περισσότερες ημέρες το τουριστικό τους φορτίο, γεγονός που προφανώς σχετίζεται με τις μεγαλύτερες δυνατότητες τουριστικής μετακίνησης που παρέχει η έκταση του οδικού δικτύου.

- *Ακαθάριστο εγχώριο προϊόν* (**x**=*GDP*): Ο συντελεστής $r_{GDP,LENGTH}$ προκύπτει σημαντικός μόνο για τη μη-μητροπολιτική περίπτωση ($r_{GDP,LENGTH}^{n_2=49} = 0,667$), καθώς για τη μητροπολιτική περίπτωση το επίπεδο σημαντικότητας αμβλύνεται στο 0.1 και η τιμή του συντελεστή μειώνεται θεαματικά ($r_{GDP,LENGTH}^{n_1=51} = 0,239$). Το θετικό πρόσημο του συντελεστή εκφράζει ότι οι νομοί με μεγάλο μήκος οδικού δικτύου τείνουν να συνεισφέρουν περισσότερο στη διαμόρφωση του ΑΕΠ της χώρας, συνδέοντας την έκταση των οδικών υποδομών με την οικονομική δυναμική των περιφερειών. Περαιτέρω, η υψηλή τιμή του συντελεστή για τη μη-μητροπολιτική περίπτωση ($r_{GDP,LENGTH}^{n_2=49} = 0,667$) υποδηλώνει ότι η οικονομική ευημερία των νομών της επαρχίας σχετίζεται σε μεγάλο βαθμό με τις οδικές τους υποδομές, γεγονός που συνηγορεί στην άποψη ότι η περιφερειακή ανάπτυξη στην Ελλάδα βασίζεται σε ένα παρωχημένο αναπτυξιακό μοντέλο.

- *Συμμετοχή του τουρισμού στο ΑΕΠ* (**x**=$T_{GDP}$): Αντίστοιχη κατάσταση με την προηγούμενη εμφανίζεται και στην περίπτωση του συντελεστή $r_{T_{GDP},LENGTH}$, αλλά αυτή τη φορά η μείωση της τιμής του από τη μη-μητροπολιτική ($r_{T_{GDP},LENGTH}^{n_2=49} = 0,301$) στη μητροπολιτική περίπτωση ($r_{T_{GDP},LENGTH}^{n_1=51} = 0,229$) προκύπτει ιδιαίτερα μικρή. Το θετικό πρόσημο του συντελεστή εκφράζει ότι οι νομοί με μεγάλο μήκος οδικού δικτύου τείνουν να έχουν υψηλό προϊόν σε τουρισμό, επαληθεύοντας τη σχέση αλληλεπίδρασης μεταξύ του τομέα των μεταφορών και του τουρισμού.

- *Πληθυσμός* (**x**=*POP*): Η εικόνα που παρουσιάζει ο συντελεστής $r_{POP,LENGTH}$ είναι αντίστοιχη με αυτή του συντελεστή $r_{GDP,LENGTH}$, δηλαδή μεγάλη και στατιστικά σημαντική τιμή για τη μη-μητροπολιτική περίπτωση ($r_{POP,LENGTH}^{n_2=49} = 0,713$), ενώ μικρή τιμή με αμβλυμμένη σημαντικότητα (σε επίπεδο σημαντικότητας 0.1) για τη μητροπολιτική ($r_{POP,LENGTH}^{n_1=51} = 0,243$). Το θετικό πρόσημο του συντελεστή εκφράζει γενικά ότι οι νομοί με μεγάλο μήκος οδικού δικτύου τείνουν να είναι περισσότερο πυκνοκατοικημένοι, κατάσταση η οποία φαίνεται πως αποτελεί τον κανόνα για τους νομούς της επαρχίας (μη-μητροπολιτική περίπτωση), περιγράφοντας ένα βαρυτικό πρότυπο (gravity model) στον αναπτυξιακό μηχανισμό των οδικών μεταφορών της χώρας.

Από την άλλη μεριά, η ανάλυση συσχετίσεων που αφορά τη μεταβλητή της πυκνότητας του εθνικού οδικού δικτύου (**y**=DENSITY) παρέχει διαφορετική πληροφορία. Με βάση τα αποτελέσματα που παρουσιάζονται στον πίνακα 4 προκύπτουν οι παρακάτω παρατηρήσεις:

- Οι συσχετίσεις μεταξύ $r_{LENGTH,DENSITY}>0$ και $r_{AREA,DENSITY}<0$ οφείλονται στον ορισμό της μεταβλητής DENSITY=(LENGTH/AREA), αλλά δεν προκύπτουν σημαντικές, ξεπερνώντας οριακά τα προεπιλεγμένα επίπεδα σημαντικότητας.

- Για τη συσχέτιση $r_{PORTS,DENSITY}$ παρατηρείται ότι αυτή στη μητροπολιτική περίπτωση είναι οριακά σημαντική σε επίπεδο σημαντικότητας $a$=0.5, έχοντας τιμή $r_{PORTS,DENSITY}^{n_1=51} = 0.,29$, ενώ στη μη μητροπολιτική περίπτωση προκύπτει στατιστικά ασήμαντη, αλλάζοντας ταυτόχρονα και το πρόσημό της ($r_{PORTS,DENSITY}^{n_2=49} = -0,201$). Η κατάσταση αυτή υποδηλώνει ότι η παρουσία των μητροπολιτικών νομών (Αττικής και Θεσσαλονίκης) ευθύνεται για την αλλαγή του προσήμου του συντελεστή συσχέτισης από αρνητικό σε θετικό, με αποτέλεσμα να παρέχονται ενδείξεις ότι οι δύο αυτοί νομοί καρπώνονται το σύνολο της θετικής συσχέτισης. Αυτό σημαίνει ότι ενώ στην περίπτωση των επαρχιακών νομών υφίστανται ενδείξεις πως η πυκνότητα του οδικού δικτύου δρα ανταγωνιστικά προς την ακτοπλοϊκή δραστηριότητα του νομού, στο βαθμό που αυτή αντιπροσωπεύεται από τον αριθμό των λιμένων, στην περίπτωση των δύο μητροπολιτικών νομών παύει να υφίσταται αυτή η διαπίστωση και δράση των δύο τρόπων μεταφοράς φαίνεται πως αντιστρέφεται από ανταγωνιστική σε συνεργατική.

- Η συσχέτιση $r_{IPP,DENSITY}$ παρουσιάζει αντίθετο πρόσημο από την αντίστοιχη $r_{IPP,LENGTH}$ και παύει να είναι στατιστικά σημαντική, όπως η δεύτερη. Το γεγονός αυτό υποδηλώνει ότι οι νομοί με πυκνό οδικό δίκτυο στην επιφάνειά τους φαίνεται πως αίρουν το πλεονέκτημα να έχουν ευκολότερη πρόσβαση στις οικονομικές δραστηριότητες των υπολοίπων. Η διαπίστωση αυτή συμπληρώνεται από τα επόμενα αποτελέσματα του συντελεστή συσχέτισης $r_{DPP,DENSITY}$, τα οποία αυτή τη φορά είναι θετικά. Ιδιαίτερα στη μητροπολιτική περίπτωση, η τιμή του συντελεστή προκύπτει αρκετά υψηλή και στατιστικά σημαντική. Το γεγονός αυτό εκφράζει ότι νομοί με πυκνό οδικό δίκτυο στην επιφάνειά τους φαίνεται πως συγκεντρώνουν τις οικονομικές δραστηριότητες και τη δυνατότητα πρόσβασης στο εσωτερικό τους, με αποτέλεσμα να περιορίζουν την πρόσβασή τους στις οικονομικές δραστηριότητες των υπολοίπων νομών, όπως φάνηκε στις τιμές των συσχετίσεων $r_{\text{IPP,DENSITY}}$. Ιδιαίτερα, οι δύο μητροπολιτικοί νομοί της Αττικής και της Θεσσαλονίκης εμφανίζονται ότι καρπώνονται το μεγαλύτερο ποσοστό της στατιστικά σημαντικής υψηλής τιμής του συντελεστή $r_{DPP,DENSITY}^{n_1=51} = 0,707$, περιγράφοντας ένα βαρυτικό πρότυπο μεγάλης συνοχής στο εσωτερικό τους.

- Παρόμοια κατάσταση με τις συσχετίσεις $r_{\text{DPP,DENSITY}}$ φαίνεται πως περιγράφει και τις συσχετίσεις $r_{\text{GDP,DENSITY}}$. Ειδικότερα, η μητροπολιτική περίπτωση εμφανίζει υψηλή και στατιστικά σημαντική τιμή του συντελεστή, γεγονός το οποίο, σε συνδυασμό με την ασήμαντη τιμή του συντελεστή για τη μη-μητροπολιτική περίπτωση, εκφράζει ότι το πυκνό οδικό δίκτυο λειτουργεί ως παράγοντας οικονομικής ανάπτυξης (στο βαθμό που η οικονομική ανάπτυξη αντανακλάται στο GDP μιας περιφέρειας) για την περίπτωση των μητροπολιτικών νομών της Αττικής και της Θεσσαλονίκης.

- Η συσχέτιση $r_{A_{SEC},DENSITY}$ παρουσιάζει αρνητικό πρόσημο και στατιστική σημαντικότητα μόνο για τη μητροπολιτική περίπτωση, ερμηνεύοντας ότι οι νομοί με πυκνό οδικό δίκτυο (με καθοριστική τη θέση των δύο μητροπολιτικών) τείνουν να εμφανίζουν χαμηλή συμμετοχή του πρωτογενή τομέα στη διαμόρφωση του ΑΕΠ της χώρας και πιο ελεύθερα χαμηλή αγροτική παραγωγή ή δραστηριότητα.

- Αντίθετα με την προηγούμενη κατάσταση, η συσχέτιση $r_{C_{SEC},DENSITY}$ είναι στατιστικά σημαντική και παρουσιάζει θετικό πρόσημο. Το γεγονός αυτό εκφράζει ότι οι νομοί με πυκνό οδικό δίκτυο τείνουν να εμφανίζουν υψηλή συμμετοχή του τομέα των υπηρεσιών στη διαμόρφωση του ΑΕΠ της χώρας. Η εικόνα αυτή φαίνεται πως εξειδικεύεται στο τουριστικό προϊόν με τη συσχέτιση $r_{T_{GDP},DENSITY}$, η οποία ακολουθεί το ίδιο μοτίβο με τη συσχέτιση $r_{C_{SEC},DENSITY}$, με μόνη διαφορά ότι στη μητροπολιτική περίπτωση ($n_1$=51) η τιμή του συντελεστή προκύπτει σχεδόν διπλάσια της μη-μητροπολιτικής. Αυτό σημαίνει ότι η πυκνότητα του οδικού δικτύου (μήκος οδών ανά νομό/επιφάνεια νομού) αποτελεί παράγοντα εξυπηρέτησης του τουρισμού και ιδιαίτερα για τις περιπτώσεις των νομών της Αθήνας και της Θεσσαλονίκης.

- Ακολούθως, η συσχέτιση $r_{AGR_{INV},DENSITY}$ είναι αρνητική και εμφανίζεται στατιστικά σημαντική μόνο για την μη-μητροπολιτική περίπτωση. Το αποτέλεσμα αυτό εκφράζει ότι οι επαρχιακοί νομοί με πυκνό οδικό δίκτυο τείνουν να γίνονται αποδέκτες χαμηλότερων επενδύσεων στην αγροτική βιομηχανία, σε σχέση με τους αραιότερους σε οδικές υποδομές. Μία περαιτέρω ερμηνεία που μπορεί να δοθεί στη συσχέτιση μεταξύ πυκνότητας οδικού δικτύου και επενδύσεων στην αγροτική βιομηχανία είναι ότι η ανάπτυξη της αγροτικής παραγωγής προϋποθέτει την ύπαρξη μεγάλων καλλιεργήσιμων εκτάσεων, περιορίζοντας έτσι τις διαθέσιμες επιφάνειες για ανέγερση οδικών υποδομών. Ωστόσο, μέσα σε αυτή την τιμή του συντελεστή ενδεχομένως να κρύβεται και η πληροφορία ότι η αγροτική παραγωγή δεν υποστηρίζεται όπως θα έπρεπε από τις οδικές υποδομές της χώρας, αναδεικνύοντας ένα θέμα για το σχετικό διάλογο.

- Παρόμοια εικόνα με τις συσχετίσεις $r_{\text{DPP,DENSITY}}$ και $r_{\text{GDP,DENSITY}}$ φαίνεται πως περιγράφει και τη συσχέτιση $r_{\text{POP,DENSITY}}$, στην οποία παρατηρείται θετικό πρόσημο, αλλά μόνο η μητροπολιτική περίπτωση εμφανίζει υψηλή και στατιστικά σημαντική τιμή του συντελεστή. Το γεγονός αυτό εκφράζει ότι οι νομοί με πυκνό οδικό δίκτυο τείνουν να είναι περισσότερο πυκνοκατοικημένοι, με καθοριστική την παρουσία των δύο μητροπολιτικών, ενισχύοντας τη βαρυτική διαμόρφωση της δομής των οδικών υποδομών.

- Στη συνέχεια, οι συσχετίσεις $r_{\text{WELF,DENSITY}}$ και $r_{\text{EDU,DENSITY}}$ εμφανίζουν παρόμοια συμπεριφορά, περιγράφοντας ότι οι νομοί με πυκνό οδικό δίκτυο τείνουν να διακρίνονται από μεγαλύτερο δείκτη ευημερίας και από υψηλότερο επίπεδο εκπαίδευσης του πληθυσμού τους. Εδώ η παρουσία των μητροπολιτικών περιπτώσεων δεν φαίνεται ιδιαίτερα καθοριστική, με αποτέλεσμα η συνολική εικόνα που διαμορφώνεται από το σύνολο των συσχετίσεων να παραπέμπει στο πρώιμο και παρωχημένο πρότυπο της *«ανάπτυξης εκεί που υπάρχει δρόμος»*.

- Η συμπεριφορά της μεταβλητής URB παράγει τόσο κατά τη συσχέτισή της με τη μεταβλητή LENGTH όσο και με τη μεταβλητή DENSITY αποτελέσματα που είναι στατιστικά ασήμαντα. Παρόλα αυτά, μόνο η τιμή του συντελεστή συσχέτισης $r_{URB,DENSITY}$ στη μη-μητροπολιτική περίπτωση έχει θετικό πρόσημο, γεγονός το οποίο εκφράζει (με την επιφύλαξη των τιμών της στατιστικής σημαντικότητας) ότι η πυκνότητα του οδικού δικτύου και η βαθμός αστικοποίησης φαίνεται πως σχετίζεται μόνο στην περίπτωση των επαρχιακών νομών.

- Τέλος, η περίπτωση των συσχετίσεων της μεταβλητής DENSITY με τις μεταβλητές τουρισμού (R, $R_T$, $R_{ST}$) παρουσιάζει αντιστραμμένη εικόνα σε σχέση με τις αντίστοιχες περιπτώσεις συσχετίσεων της μεταβλητής LENGTH. Ειδικότερα, ο συντελεστής $r_{R,DENSITY}$ εμφανίζεται θετικός και με υψηλότερη τιμή στην περίπτωση $n_2=49$, γεγονός το οποίο υποδηλώνει ότι οι νομοί με μεγάλη πυκνότητα οδικού δικτύου και ιδιαίτερα αυτοί της επαρχίας τείνουν να αναλαμβάνουν μεγαλύτερο φορτίο τουρισμού. Περαιτέρω, οι συσχετίσεις $r_{R_T,DENSITY}$ εμφανίζουν αρνητικό πρόσημο και, με την επιφύλαξη της μη στατιστικά σημαντικής τιμής τους, υπονοούν ότι οι νομοί με μεγάλη πυκνότητα οδικού δικτύου τείνουν να δέχονται μικρότερο αριθμό τουριστών. Τέλος, η περίπτωση της συσχέτισης $r_{R_{ST},DENSITY}$ παρουσιάζεται σε πλήρη συμφωνία με τις αντίστοιχες συσχετίσεις ως προς τη μεταβλητή LENGTH, εμφανίζοντας υψηλότερες τιμές από τις τελευταίες αναφερόμενες. Το γεγονός αυτό υποδηλώνει ότι οι νομοί με μεγάλη πυκνότητα οδικών υποδομών τείνουν να διατηρούν για περισσότερες ημέρες το τουριστικό τους φορτίο.

## 4. Συμπεράσματα

Στο άρθρο αυτό μελετήθηκε η τοπολογία του διαπεριφερειακού οδικού δικτύου της Ελλάδας (Greek Road Network - GRN), με χρήση της ανάλυσης σύνθετων δικτύων (complex network analysis) και στατιστικής μηχανικής. Σκοπό της μελέτης αποτέλεσε η εξόρυξη της κοινωνικοοικονομικής πληροφορίας που είναι ενσωματωμένη στην τοπολογία του GRN, προκειμένου να διερευνηθούν οι παράγοντες του δικτύου που σχετίζονται επιδρούν στην περαιτέρω οικονομική και περιφερειακή ανάπτυξη της χώρας. Το GRN αναπαραστάθηκε στον *L*-χώρο αντιπροσώπευσης ως μη κατευθυνόμενος γράφος, όπου το σύνολο των κόμβων *V* αντιστοιχεί σε *διασταυρώσεις (intersections)* διαδρομών, ενώ το σύνολο των ακμών *E* σε *διαδρομές μονής διεύθυνσης* (δίχως αλλαγή πορείας).

Στην ανάλυση που πραγματοποιήθηκε διαφάνηκε ότι στην τοπολογία του GRN καθίσταται εμφανής η επίδραση των χωρικών περιορισμών, όπως στοιχειοθετείται από τα οξυμένα πρότυπα στην κατανομή του βαθμού, τη σημαντική συγκέντρωση των τιμών γύρω από την κύρια διαγώνιο στα πρότυπα σποραδικότητας (sparsity patterns – spy plots) των πινάκων συνδέσεων, τον υπολογισμό του *ω*-δείκτη, ο οποίος χρησιμοποιείται για την προσεγγιστική ανίχνευση τυπολογίας δικτύων, τις χωρικές κατανομές των βασικών μέτρων και μεγεθών δικτύων (βαθμός, ενδιαμεσότητα, εγγύτητα, συγκέντρωση, συναρμολογησιμότητα, χωρική ισχύς), τον επιμερισμό του δικτύου σε κοινότητες γεωγραφικής συνάφειας, τα πρότυπα *κανόνα-δύναμης (power-law)* που διαμορφώθηκαν από τις συσχετίσεις (*k*, $C^b$), (*k*, *s*) και (*k*, *C*), και τις μεγάλες διακυμάνσεις που εμφανίστηκαν στο μέγεθος της ενδιαμέσου κεντρικότητας ($C^b$), υποδηλώνοντας τη σαφή γεωγραφική υπόσταση του δικτύου. Συνολικά, η τοπολογία του GRN διαφάνηκε συγγενική με το θεωρητικό (μηδενικό) πρότυπο του *δικτυώματος* (lattice network), η οποία σχετίζεται με την ύπαρξη χωρικών περιορισμών.

Επίσης, μελετήθηκε εμπειρικά το ποσό της κοινωνικοοικονομικής πληροφορίας που περιέχεται στην τοπολογία του GRN, με χρήση ανάλυσης συσχετίσεων μεταξύ ενός συνόλου χωρικών, οικονομικών, δημογραφικών και τουριστικής προέλευσης μεταβλητών και καθεμιάς από τις δύο βασικές μεταβλητές υποδομής του οδικού δικτύου (μήκος και πυκνότητα), οι οποίες διαμορφώθηκαν με σημείο αναφοράς το νομό. Για την αξιολόγηση της επιρροής που έχουν οι νομοί της Αθήνας και της Θεσσαλονίκης (μητροπολιτικές περιπτώσεις) στη διαμόρφωση των παραπάνω συσχετίσεων, θεωρήθηκαν δύο περιπτώσεις συνόλων, η μία συμπεριλαμβανομένων και η δεύτερη εξαιρουμένων των δύο μητροπόλεων. Οι υπολογισμοί των συντελεστών για τη μη-μητροπολιτική περίπτωση και ως προς τη μεταβλητή του μήκος του οδικού δικτύου διαφάνηκε ότι οδηγούν κατά κανόνα στην παραγωγή ευνοϊκότερων αποτελεσμάτων από τη μητροπολιτική περίπτωση, ενώ αυτοί που πραγματοποιήθηκαν ως προς τη μεταβλητή της πυκνότητας εμφάνισαν σημαντικές αναντιστοιχίες, δίχως να οδηγούν σε κάποιον γενικό κανόνα.

Η ανάλυση συσχετίσεων παρείχε γενικά ενδείξεις ότι ο μηχανισμός της χωρικής κατανομής των οδικών υποδομών περιγράφεται από *βαρυτική λειτουργία*, έχοντας δηλαδή ως προτεραιότητα την εξυπηρέτηση των περιοχών ανάλογα με τον πληθυσμό τους, αλλά και από το βασικό και ίσως παρωχημένο πρότυπο της «*ανάπτυξης εκεί που υπάρχει δρόμος*», το οποίο διέπνεε τη νοοτροπία των προηγούμενων δεκαετιών στην Ελλάδα. Η διαπίστωση περί παρωχημένου αναπτυξιακού προτύπου βασίστηκε στην παρατήρηση ότι οι τιμές των συντελεστών συσχέτισης που περιγράφουν γενικά το ζεύγος «*ευημερία – οδικές υποδομές*» εμφανίστηκαν υψηλότερες για την περίπτωση *των επαρχιακών νομών* (μη-μητροπολιτική περίπτωση). Οι νομοί με μεγάλο μήκος οδικού δικτύου διαφάνηκε ότι τείνουν να είναι περισσότερο πυκνοκατοικημένοι (ιδιαίτερα για τους νομούς της επαρχίας), να έχουν μεγαλύτερο συγκριτικό πλεονέκτημα για οικονομική ανάπτυξη, σχετικά υψηλότερο επίπεδο εκπαίδευσης του πληθυσμού τους, καλύτερη δυνατότητα εξυπηρέτησης της τουριστικής μετακίνησης, καλύτερο επίπεδο ευημερίας και μεγαλύτερη συνεισφορά στο ΑΕΠ της χώρας. Από την άλλη πλευρά, οι νομοί που διαθέτουν μεγάλο μήκος οδικού δικτύου τείνουν να μην απολαμβάνουν εξίσου μεγάλες επενδύσεις στην αγροτική παραγωγή, γεγονός το οποίο περιγράφει μία παθογένεια στη λειτουργία και στις αναπτυξιακές δυναμικές του πρωτογενή τομέα της χώρας, λαμβάνοντας υπόψη ότι ο ρόλος των οδικών υποδομών στην υποστήριξη των αγροτικών δραστηριοτήτων είναι καθοριστικός.

Τέλος, η ανάλυση συσχετίσεων που πραγματοποιήθηκε ως προς την πυκνότητα του εθνικού οδικού δικτύου συμπλήρωσε την προηγούμενη εικόνα, αναδεικνύοντας σε ορισμένες περιπτώσεις ενδιαφέρουσες αντιθέσεις μεταξύ των μητροπολιτικών και μη-μητροπολιτικών θεωρήσεων. Ο βασικός άξονας συμπερασμάτων περιγράφει και σε αυτήν την περίπτωση μία βαρυτική λειτουργία, η οποία εκφράζεται με την καθοριστική επίδραση των μητροπόλεων στην αύξηση των παρατηρούμενων συσχετίσεων για τις περιπτώσεις των μεταβλητών του πληθυσμού, του ίδιου πληθυσμιακού δυναμικού, του αριθμού των λιμένων, του εγχώριου προϊόντος, της ευημερίας και του βαθμού αστικοποίησης. Από την άλλη πλευρά, παρατηρήθηκε ότι για τους επαρχιακούς νομούς η πυκνότητα του οδικού δικτύου δρα ανταγωνιστικά προς την ακτοπλοϊκή δραστηριότητα του νομού, υποστηρίζει την υψηλή συμμετοχή του τομέα των υπηρεσιών και του τουρισμού στη διαμόρφωση του ΑΕΠ της χώρας και εμφανίζεται αρνητική στην υποστήριξη της πρωτογενούς παραγωγής (τόσο στον τομέα των επενδύσεων όσο και στη διαμόρφωση του ΑΕΠ), σκιαγραφώντας μία εικόνα υστέρησης που ενισχύει τη διαπίστωση περί παρωχημένου αναπτυξιακού προτύπου.

Συνολικά, το άρθρο αυτό αναδεικνύει την αποτελεσματικότητα της χρήσης της ανάλυσης των σύνθετων δικτύων στη μοντελοποίηση των χωρικών δικτύων και ειδικότερα των συστημάτων μεταφορών. Η εφαρμογή αυτής της σύγχρονης μεθοδολογικής προσέγγισης στον τομέα των μεταφορών αποβαίνει ιδιαίτερα αποτελεσματική, λόγω της μικτής παραμετρικής (πινακοποίηση του συστήματος που επιτρέπει τη διενέργεια υπολογισμών) και μη παραμετρικής (απεικόνιση του συστήματος σε γράφο και χαρτογράφηση της τοπολογίας του) φύσης της, παρέχοντας δυνατότητες μοντελοποίησης και περιγραφής των προβλημάτων χωρικής αλληλεπίδρασης. Για το λόγο αυτό, η παρούσα προσέγγιση επιδιώκει την προώθηση χρήσης του παραδείγματος των δικτύων στις χωρικές και περιφερειακές εφαρμογές.

## 5. Βιβλιογραφία